\documentclass[10pt,journal,compsoc]{IEEEtran}
\usepackage{amsmath,amsfonts}
\usepackage{algorithmic}
\usepackage{algorithm}
\usepackage{array}
\usepackage[caption=false,font=normalsize,labelfont=sf,textfont=sf]{subfig}
\usepackage{textcomp}
\usepackage{stfloats}
\usepackage{url}
\usepackage{verbatim}
\usepackage{graphicx}
\usepackage{cite}
\usepackage{multicol}
\usepackage{relsize}
    
\newcommand{\eg}{\textit{e.g.,~}}
\newcommand{\ie}{{\textit{i.e.},~}}
\newcommand{\vs}{\textit{vs.~}}
\newcommand{\etal}{\textit{et al}}

\usepackage[dvipsnames]{xcolor}
\definecolor{greytext}{gray}{0.5}
\definecolor{DarkGreen}{rgb}{0.0, 0.5, 0.0}
\definecolor{DarkKhaki}{rgb}{0.74, 0.72, 0.42}
\definecolor{DarkRed}{rgb}{0.7, 0.2, 0.2}
\definecolor{Purple}{rgb}{0.7,0.0,0.7}
\definecolor{Orange}{rgb}{0.7,0.2,0}
\definecolor{Teal}{rgb}{0.12,0.5,0.5}
\definecolor{Black}{rgb}{0,0,0}
\definecolor{DeepPink}{rgb}{1,0.08,0.34}

\definecolor{kmCGreen}{rgb}{0.4, 0.6, 0.6}
\definecolor{kmEGreen}{rgb}{0.0, 0.42, 0.24} 

\newcommand{\kmE}[1]{\textcolor{kmEGreen}{#1}}

\newcommand{\inlineHeading}[1]{\vspace{0.04in}\noindent {\textbf{#1}}} 

\newcommand{\lowHeading}[1]{\vspace{0.03in}\noindent\textit{#1:}}

\newcommand{\mwislong}[0]{\textit{MyWeekInSight}}
\newcommand{\mwis}[0]{\textit{MWIS}}

\newcommand{\facets}[0]{{Facets}}
\newcommand{\facet}[0]{{Facet}}

\newcommand{\goals}[0]{\textit{Requirements}}
\newcommand{\task}[0]{\textit{Task}}
\newcommand{\tasks}[0]{\textit{Tasks}}

\newcommand{\oldNumber}[1]{}

\newcommand{\chartname}[1]{\textit{#1}}    
\newcommand{\protoname}[1]{\texttt{#1}}

\definecolor{WorryBlue}{rgb}{.49, .75, .84} 
\definecolor{SleepGreen}{rgb}{0, .69, .26}
\definecolor{PeerPurple}{rgb}{.75, .62, .87}  
\definecolor{SymptomsRed}{rgb}{1.0, .45, .35}

\newcommand{\styleWorry}[1]{\colorbox{WorryBlue}{#1}}    
\newcommand{\styleSleep}[1]{\colorbox{SleepGreen}{#1}}    
\newcommand{\stylePeer}[1]{\colorbox{PeerPurple}{#1}}    
\newcommand{\styleSymptoms}[1]{\colorbox{SymptomsRed}{#1}}

\makeatletter
\def\testclr#1#{\@testclr{#1}}
\def\@testclr#1#2{{\fboxsep\z@\fbox{\colorbox#1{#2}{\phantom{XX}}}}}

\makeatother

\usepackage{pdfpages}
\usepackage{booktabs}  

\usepackage{enumitem}
\setitemize{itemsep=1pt,topsep=2pt,parsep=0pt,partopsep=0pt}

\newcommand{\suppRef}[1]{{\S}Sup.{#1}}   

\newcommand{\SuppTaskAbs}[0]{\suppRef{1}}
\newcommand{\SuppVizChanges}[0]{\suppRef{2}}
\newcommand{\SuppVizFinal}[0]{\suppRef{3}}
\newcommand{\SuppVizRedesign}[0]{\suppRef{4}}
\newcommand{\SuppEMASurvey}[0]{\suppRef{5}}
\newcommand{\SuppDesignEval}[0]{\suppRef{6}}
\newcommand{\SuppImpactEval}[0]{\suppRef{7}}
\newcommand{\SuppExamples}[0]{\suppRef{8}}
\newcommand{\SuppQuant}[0]{\suppRef{9}}

\begin{document}

\title{\mwislong: \\ \smaller{Designing and Evaluating the Use of Visualization \\ in  Self-Management of Chronic Pain by Youth}}




\author{Unma Desai, Haley Foladare, Katelynn E. Boerner, Tim F. Oberlander, Tamara Munzner, Karon E. MacLean
  \IEEEcompsocitemizethanks{
    \IEEEcompsocthanksitem Unma Desai, Haley Foladare, Tamara Munzner, and Karon E. MacLean are with the University of British Columbia Department of Computer Science. E-mail: \{unma, foladare, tmm, maclean\}@cs.ubc.ca. \protect\\
    \IEEEcompsocthanksitem Katelynn E. Boerner (Ph.D.) and Tim F. Oberlander (M.D.) are with University of British Columbia
Department of Pediatrics and the BC Children’s Hospital and Research Institute.
E-mail: \{katelynn.boerner, toberlander\}@bcchr.ca. }
  \thanks{Manuscript received May 29, 2024; revised XXX.}
}

\markboth{IEEE TRANSACTIONS ON VISUALIZATION AND COMPUTER GRAPHICS, ~Vol.~xx, No.~x, xxxx~xxxx}%
{Author \MakeLowercase{\textit{et al.}}: \mwislong: Designing and Evaluating the Use of Visualization in Self-Management of Chronic Pain by Youth}
%



\IEEEtitleabstractindextext{
\begin{abstract}
A teenager's experience of chronic pain reverberates through multiple interacting aspects of their lives. 
To self-manage their symptoms, they need to understand how factors such as their sleep, social interactions, emotions and pain intersect; supporting this capability must underlie an effective personalized healthcare solution. While adult use of personal informatics for self-management of various health factors has been studied, solutions intended for adults are rarely workable for teens, who face this complex and confusing situation with unique perspectives, skills and contexts.
In this design study, we explore a means of facilitating self-reflection by youth living with chronic pain, through visualization of their personal health data. 
In collaboration with pediatric chronic pain clinicians and a health-tech industry partner, we designed and deployed \textit{MyWeekInSight},
a visualization-based self-reflection tool for youth with chronic pain. We discuss our staged design approach with this intersectionally vulnerable population, in which we balanced reliance on proxy users and data with feedback from youth viewing their own data. We report on extensive formative and in-situ evaluation, including a three-week clinical deployment,
and present a framework of challenges and barriers faced in clinical deployment with mitigations that can aid fellow researchers. 
Our reflections on the design process yield principles, surprises, and open questions.
\end{abstract}

}

\maketitle

\IEEEdisplaynontitleabstractindextext

\IEEEraisesectionheading{
\section{Introduction}
	\label{Sec:Intro}}	

\IEEEPARstart {C}{hronic} pain is a common and costly condition, 
affecting 10\% of the population worldwide~\cite{jackson_global_2014}. 
In this design study, we explore the potential of visualization of personal data for the self-management of chronic pain by adolescents. 
Although the factors that underlie chronic pain have been extensively studied~\cite{stinson_chronic_2009, groenewald_economic_2014, murray_long-term_2020}, visualization in support of pain self-management in youth has not.

Adolescent patients with chronic pain are just learning to navigate shifting social dynamics, regulate intense emotions, and understand physiological needs like rest, nutrition, and exercise. 
These teen rites of passage are amplified and complicated by interacting factors of the pain itself, constraints and perceptions associated with their specific condition, and complex, time-consuming relations with an extensive care team~\cite{stinson_chronic_2009}. 
Supporting teens in self-managing their health requires a personalized and holistic perspective on their everyday lives. 
We seek to engage youth in reflecting on interconnections between aspects of their own lives -- \eg how pain co-varies with sleep~\cite{Morris_Pediatric_2022}, emotions~\cite{vinall_mental_2016}, and peer relationships~\cite{Jones_Sociodevelopmental_2020}. 
As a self-appraised condition, there is a potential for a patient's pain experience to be influenced by their predispositions and meaning-making relative to their symptoms and triggers. 
For example, increased understanding and expectations of one's symptoms obtained via daily symptom self-report could motivate a person to alter behaviour patterns associated with pain~\cite{ancker:2015:dataTrackMCC,cushing_tailoring_2019}
or support them with psychological reframing to mitigate the pain's impact~\cite{Coakley_Evidence_2017}.

We hypothesize that fine-grained, longitudinal record-keeping combined with reflection on factor covariance could scaffold such an understanding, and thus facilitate well-being, with benefits sufficiently concrete to regard visualization as a clinical intervention~\cite{boerner_data_2022}.
We explore this premise in a collaboration initiated by clinicians and patient partners at a pediatric hospital and associated research institute, who recruited human-computer interaction and visualization experts.	
In the smartphone-based visualization application we designed, youth self-report the data that will be visualized through an Ecological Momentary Assessment (EMA) to provide accurate in-the-moment data about their pain and lived experiences that are not reliant on recall~\cite{boerner_data_2022}.
We discuss design challenges encountered in communicating this personal data back to the patient in a way that enhances engagement and promotes self-understanding. 
For example, we contended with the same combination of small screens and limited attention common in mobile visualization design~\cite{lee_2021}; 
but in contrast to the many design studies addressing scalability of large data volume~\cite{lex_entourage_2013, 
	meyer_multeesum_2010,
	pretorius_visualization_2011, 
	williams_visualizing_2020}, 
we have found interesting and under-explored challenges in the presentation of data at a very small and personal scale.

We reflect on the challenges of requirements analysis, design and in particular, deployment for a population with the intersectional vulnerabilities of clinical patient and teenaged minor. Our clinician team members provided extensive feedback throughout design, deployment, and evaluation. Restricted access to and complexity of working with this population led us to a staged design and evaluation~\cite{mclachlan_2008}, with judicious enlistment of proxy users and recourse to proxy data in two early studies to conserve the limited resource of youth patients' time for our final two studies.

\vspace{0.02in} \noindent
We contribute:

\inlineHeading{1. Data abstractions, requirements, and task abstractions} for using visualizations as a treatment intervention for youth with chronic pain.
        
\inlineHeading{2. The design and development of \mwislong\ (\mwis),} a visualization application enabling self-reflection via self-gathered personal data collected over a single week, developed with continuous clinician involvement. 
We assessed \mwis\ through 
        a usability pilot (N=6) with proxy users and data, a preliminary utility study (N=10) with real users and proxy data, and a three-week clinical deployment (N=44) with real users and real data in conjunction with a utility study with a subset of these users (N=11).

\inlineHeading{3. Reflections on deployment} of digital applications as treatment interventions with intersectionally vulnerable populations, with insights structured as challenges, barriers, and mitigations.

\inlineHeading{4. Reflections on design principles, surprises, and open questions} that emerged from our process.

\section{Related Work}
	\label{Sec:RelatedWork} 

Our work lies at the intersection of data tracking for chronic pain or self-quantification 
and of visualization of personal health data, both contextualized for teen patients. We further situate our work with respect to currently deployed applications for pain self-management. 

\subsection{Chronic Pain and Pain-Tracking Methods}
	\label{Subsec:RW:CPTracking}		

Defined as pain that lasts over three months, chronic pain may or may not have a known underlying cause~\cite{mills_chronic_2019}. In youth aged 19 and under, it is a common 
condition; 3-5\% of sufferers report disabling pain levels~\cite{king_epidemiology_2011}. It has many potential causes, exacerbated by social, psychological, physiological and environmental parameters like age, gender, stress, family attitudes and social interactions~\cite{stinson_chronic_2009}.
Pain can interfere with academic, social, and recreational functioning, mental health, and quality of life~\cite{groenewald_economic_2014, murray_long-term_2020,noel_chronic_2016, vinall_mental_2016, dewar_using_2003}.

Diagnosis and treatment of pain relies on the patient's recollection of pain episodes, and levels of pain within them. Traditional recall tools like journals often introduce and exacerbate bias due to delay in recording~\cite{van_den_brink_occurrence_2001, stone_patient_2002}. 
Real-time data capture methods like Ecological Momentary Assessment (EMA)~\cite{shiffman_ecological_2008}, where participants are repeatedly prompted to report their current behavior and experiences grounded in their lived environment, can help them capture accurate in-the-moment data with minimal intrusion.

\subsection{Quantified Self Tracking}
	\label{Subsec:RW:QuantSelf}	

\textit{Quantified Selfers} (or Q-Selfers) are individuals or groups who track personal health data (like diet, sleep, exercise) with motives of gaining health insights or managing health conditions~\cite{swan_2013}.
In a survey, Choe \etal~\cite{choe_understanding_2014} 
found that leading motivation for Q-Selfers was to improve health, specifically by managing a condition or identifying relationships between factors. 
Today, there is extensive support for digital self-tracking, often with data-collecting wearables.  

\inlineHeading{General Benefits of Self-Tracking:}
Self-tracking personal health information has been linked to positive health outcome impacts for both adults and youth. 
Stiglbauer \etal~\cite{stiglbauer_does_2019} 
found small but significant positive effects on users' self-reported health and well-being following two weeks of wearing a self-tracking device. They also found correlations between engagement and well-being, as users who engaged more with the accompanying mobile app reported stronger increases in self-reported physical health, positive emotions, and experienced accomplishment.
Riggare \etal~\cite{riggare_you_2019} surveyed 280 members of the Swedish Parkinson's Disease community for their opinions on self-tracking, and proposed the use of self-tracking as a personal improvement project to understand and improve one's health. Responses stated using self-tracking to understand how different factors affected their experience with the disease, including valuable insights gained from tracking medication type, intake, and physical activities. 

\inlineHeading{Teenagers and Self-Tracking:}
Freeman \etal~\cite{lee_freeman_ask_2022} conducted co-design workshops with 44 teenagers to find that teenagers understood and wanted to be cautious of obsessive self-tracking, preferring supportive language rather than a motivation-based, behaviour-change-oriented one in self-tracking applications. They considered privacy and autonomy important, to have control over their data tracking and sharing practices.
Stinson \etal~\cite{stinson_icancope_nodate} conducted interviews and focus groups with 23 adolescents to design a self-management application for youth with chronic pain, where they found adolescents willing and interested in tracking their symptoms to recognize patterns.  
In contrast to adult populations, Potapov \etal~\cite{potapov_what_2021} found adolescents were also more motivated by social and emotional factors, such as connecting with friends and receiving feedback, than by health-related factors. 
Freeman \etal~\cite{freeman_challenge_2021} further elaborate on the nuanced differences of self-tracking for adults \vs adolescents, considering factors such as institutional constraints and adolescents' changing physical bodies and social pressures, and highlight the need for adolescent-centered self-tracking tools.
In a qualitative study with 166 youth aged 16-18 years, they found that adolescents preferred self-tracking applications tailored to their specific needs, and which provided holistic rather than goal-driven personalized feedback and support. 

In \mwis, we prioritized holistic reflection (\ie considering cross-coupling of diverse life dimensions), leaving personalization for a future point when we have more data on where it is most needed.

\subsection{Personal Health Visualizations}
	\label{Subsec:RW:PersonalVis}	

Personal health information plays a major role in the burgeoning field of personal visualizations and their analytics~\cite{huang_2015}. 
Kim \etal~\cite{kim_dataselfie_2019} found that personal visualizations can be a valuable tool for self-reflection and behavior change. They also state that personal data, when collected qualitatively, could allow for a more nuanced and richer reflection than is possible with current automated tracking processes.  

\inlineHeading{Diverse Approaches to Interacting with Personal Data:}
People choose to interact and explore their personal health data in different ways with different aims. For instance, Thudt \etal~\cite{thudt_self-reflection_2018} discuss how insights from personal health data can be data-driven (derived directly from data) or action-driven (due to the act of recording data or seeing the visualizations). 
In a multistage longitudinal study, Moore \etal~\cite{moore_exploring_2022} found that people interacted with their data in different ways; some were goal-driven, while others interacted in playful ways, but discovered insights through exploration. 

{Short-term and unconventional} self-tracking can lead to effective personal visualizations,
as with the \textit{Dear Data} project's showcasing of hand-drawn custom visualizations to track and gain insights into daily behaviours~\cite{lupi_2016}.
	
\inlineHeading{Common Personal Data Visualization Formats:}
Choe \etal\ examined how visualization systems can help users draw insights from their own data, by studying 30 presentation videos from a meetup where Q-selfers presented their self-tracking experiences~\cite{choe_characterizing_2015}. 
This group tended to find their insights via self-reflection based on trends, comparisons or correlations of recorded data, and typically used basic visualization forms such as line charts, barcharts, and scatterplots, created both in user-customized spreadsheets and common health tracking applications such as Fitbit~\cite{fitbit} and Apple Health~\cite{noauthor_ios_nodate}, which visualize data such as sleep and activity levels. 
Choe \etal\ therefore advised designing personal data visualizations to prioritize self-reflection as a path to behaviour change, and to rely on basic visual forms. 

We found a similar user preference for simple chart forms, and in \mwis\ primarily leveraged line and barcharts arranged to promote comparison and self-reflection.

\inlineHeading{Visualizations in a Clinical Context:}
Complementing self-reflection, in a patient-centric context, data can be shared with clinicians to collaborate on treatment plans.
In an interview study (N=21), Zhu \etal~\cite{zhu_sharing_2016} found that patient-initiated data gathering, and often visualization, was motivated by desires for self-awareness and self-management. Their data and visualizations also helped them better collaborate with and answer their clinicians. Conversely, clinicians mentioned inaccuracies or clinically irrelevant data  as reasons they could not trust patient-generated data, for which a solution could be clinically-guided secure applications. 

In \mwis' design, in every step from design to deployment we involved clinicians experienced with the target population, relying on their knowledge of both clinical and teen-interest relevance for patient-generated tracking choices and to avoid common causes of data inaccuracy.

\inlineHeading{Visualizations in an Adolescent Context:}
Despite extensive guidelines for general visualization design~\cite{munzner_2014}, there has been minimal attention to designing visualizations aimed at teens. 
One exception evaluates the ability of teens to comprehend visualizations for improving soccer prowess, noting different ability levels in comprehending personal visualization data~\cite{herdal_designing_2016}.

\subsection{Pain Self-Management Applications}
	\label{Subsec:RW:CurrentApps}	

Pain management applications abound in the commercial market, with over 431 available on Android and iOS~\cite{macpherson_pain_2022} at this time. Some, \eg manageMyPain~\cite{manage_my_pain} and migraineBuddy~\cite{migraine_buddy}, help track certain symptoms for adult patients with chronic pain. 
Most of these commercially available applications rely on users to input data on a periodic basis, and use some form of gamification or visualization to show patterns in their data to users in hopes of stimulating engagement. Very few are clinically validated or even tested with a clinical population of target end users~\cite{macpherson_pain_2022, lalloo_theres_2015}.

Many applications available online do not adequately address user requirements, leading to sub-optimal engagement from the target users~\cite{rahman_managemypain_engagement_2017, eysenbach_law_2005}. Safi \etal~\cite{safi_empirical_2019} analyzed survey results from over 9000 users of healthcare apps, noting that more than 50\% of users stopped using the apps after a few weeks, despite finding value in them. 
Oakley-Girvan \etal~\cite{oakley-girvan_what_2022} also reported high drop-offs after the first week for mobile app interventions, stating the most engaging mobile interventions were ones that provided participants with the ability to view and/or interact with their health data.

Some non-commercial applications have been created and validated by clinical research teams. 
Stinson \etal's e-Ouch~\cite{stinson_e-ouch_2006} used a similar EMA approach to gather data as \mwis, but was not tested in the participants' natural environments, and did not include visualizations of the patient data. 
Jibb \etal~\cite{jibb_implementation_2017} created a web-based smartphone application with a 22-item questionnaire to assess adolescent cancer pain each morning and evening and tested it with 40 adolescents. They found significant improvements in social and emotional (although not physical and school) functioning, showing the potential of real-time pain management through smartphone applications. 

Attempts to bridge the gap between commercial availability and clinically informed design are rare. A notable exception is a gamification-based app for adults with chronic pain created in collaboration with 11 health professionals and 2 digital health experts, from Hoffmann \etal\cite{hoffmann_toward_2020}. 
They reported positive feedback for the app's potential but had not tested it with their target demographic of patients with chronic pain. 
Cooke \etal~\cite{cooke_mypainpal_2021} created myPainPal, an app for managing chronic pain in young people, that incorporates feedback from patients, parents, and clinicians, as well as a pain advisory group. The app's monitoring and tracking aspects received positive feedback from patients and parents. However, myPainPal serves primarily as an in-the-moment reflective tool, with features like a daily diary and goal setting; it does not reflect data to the user over time. 

\mwis\ approaches the same goal of promoting self-reflection by  displaying patient-provided data through engaging visualizations in a way that invites connections between life dimensions, and brings in additional factors like mental health and social interactions. 

Overall, we have an incomplete picture of compliance and engagement with chronic pain self-management tools due to a lack of longitudinal, daily-life testing.
A rare exception from Mamykina \etal~\cite{mamykina_scaling_2021}  discusses challenges with deploying long-term (3-4 years) clinical self-management interventions in the wild, providing rich insights on technological aspects and user engagement changes over time.

\section{Process}
	\label{Sec:Process}	

\begin{figure*}
    \centering
    \includegraphics[width=\textwidth]{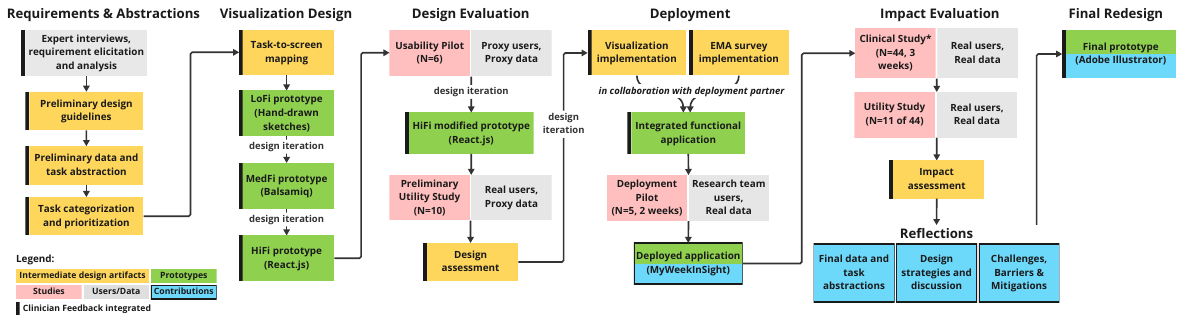}
    \caption{\mwis's seven-stage design, deployment and evaluation process. 
    \textit{*Clinical study design~\cite{boerner_data_2022} and results~\cite{boerner_making_2023} reported elsewhere}. 
    }

    \label{fig:process}
\end{figure*}
 
Our team consisted of four human-computer interaction (HCI) and visualization experts (referred to as the \textbf{design team} throughout) and two clinician-researcher domain experts (\textbf{clinicians}); all are co-authors. The clinicians had long worked with youth with chronic pain and previously created the EMA survey instrument to collect data from these patients~\cite{boerner_data_2022}. 
We slightly refined the EMA survey over the course of this project to adapt to clinical and technical needs, ultimately using the version shared in \SuppEMASurvey.

In this design study we mostly adhered to the methodology of Sedlmair \etal\cite{sedlmair_2012}, including iterative refinement of both abstractions and designs, and close collaboration between the design team and clinicians with weekly meetings at all project stages. We also incorporated a staged methodology~\cite{mclachlan_2008}, since the time and attention of our intersectionally vulnerable population was a very limited resource to be carefully conserved. 
In early rounds of design and evaluation, we carefully chose when to substitute proxy users for target users, and when to use proxy EMA survey data generated by members of the research team rather than requiring study participants to collect their own data. 

Figure~\ref{fig:process} details the project's seven stages, carried out (with overlap) over $\sim$27 months.
Additional details on process and results 
are available in the first author's thesis~\cite{Desai_2022_thesis}.

\inlineHeading{Requirements and Abstractions     
    (\S\ref{Sec:Abstractions}; 4 months):} 
Prior to obtaining direct access to the target user population, we elicited \textbf{preliminary requirements} through expert interviews, gathering information about the target population and investigating how a visualization-based intervention might meet the needs of both patients and clinicians. 
In total we interviewed 9 
stakeholders: 2 clinician co-authors (C1 and C2), 2 young adults, 2 partial-proxy users (adults with past chronic pain experience), and 3 design experts. 

The design team then created an initial data abstraction by categorizing items in the EMA survey previously designed by the clinicians into \textbf{\facets}~representing cohesive aspects of the youths' daily lived experience. 
In parallel, thematic analysis of expert interview transcripts informed a preliminary two-level task abstraction, with overall \textbf{\goals} and the specific user \textbf{\tasks} to support.
The team then iteratively refined these preliminary data and task abstractions through online affinity mapping, with clinician feedback at weekly full-team meetings. 

Eventually, we categorized each task on this comprehensive list according to patient- or clinician-relevant focus, three task priorities (high, medium, low), and level of engagement required (reflective, non-reflective). 
To constrain scope, we eliminated low-priority tasks and gave highest weight to patient-oriented reflective tasks. 
Finally, we clustered the remaining tasks into three purposes:
    \textit{Discover extremes, 
    Investigate trends over time}, and 
    \textit{Compare attributes}.

\inlineHeading{Visualization Design (\S\ref{Sec:VizDesign}, 5 months):} 
We next mapped the chosen tasks to discrete view-sets that could fit within a single smartphone screen.
We iteratively created and refined individual views 
with weekly clinician input. We began with hand-drawn sketches to brainstorm different encodings for the data and tasks, from which emerged a low-fidelity prototype (\protoname{LoFi}). 
We revised and merged encodings into a medium-fidelity Balsamiq~\cite{noauthor_balsamiq_nodate} prototype (\protoname{MedFi}). We further iterated these designs into a high-fidelity prototype (\protoname{HiFi}, React.js) for initial evaluation of the visualization design. We reviewed this high-fidelity prototype and our design evaluation interview questions with our patient partner, and incorporated this preliminary feedback.

\inlineHeading{Design Evaluation (\S\ref{Subsec:Results:UtilPrelimStudy}, 3 months):}
We explicitly evaluated the \protoname{HiFi}
visualization design in two steps. A first \textit{usability pilot} with computer science graduate students (N=6) as proxy users was followed by a design iteration of the prototype based on feedback provided. We then undertook a \textit{preliminary utility study} with actual target users, youth with chronic pain (N=10; 7F, 3M, aged 12-18 years). Both evaluations consisted of semi-structured interviews yielding qualitative data and a Likert questionnaire with quantitative data; both focused on the understandability, potential utility, and aesthetics of the visualizations. 	
In both cases, we used proxy data to generate the visualizations shown to participants. The results of these studies informed another round of prototype redesign 
(\protoname{HiFiMod}), also using React.js.

\inlineHeading{Deployment (\S\ref{Sec:DeploymentReflect}, 5 months):} 
We required a robust and secure real-world deployment to support a longitudinal clinical study. Our deployment evaluation aimed to assess the effectiveness of the coupled EMA survey and data visualizations, on the target population of youth with chronic pain. 
Because collecting personal health data imposes strict legal standards for security and privacy, we conservatively chose to work with an existing health-tech platform partner (referred to throughout as our \textbf{HTP partner}) to deploy the EMA survey and visualizations, rather than creating a new custom platform. 
Their commercial platform consisted of a general-purpose, smartphone-friendly website (rather than a native smartphone app) with built-in capabilities including user login, survey filling, dashboard viewing and text message/email reminder functions.
	
To implement the visualizations on the HTP platform, we needed to revise some encodings from the high-fidelity prototype due to platform constraints; we also made further changes to incorporate some formative feedback from the Design Evaluation study. We finalized the clinical study design with the decision that only one week of data would be shown, for consistency across participants based on randomization, rather than two weeks' worth shown in previous prototypes.
We collaborated with the HTP development team in this re-design, resulting in the deployed version (\protoname{Deploy}), also referred to as the visualization ``\textbf{dashboard}'' below. The complete list of changes from \protoname{HiFiMod} to \protoname{Deploy} is provided in \SuppVizChanges. 

The research team piloted the EMA+visualization deployable application for two weeks, followed by a round of technical bug fixing.
The final deployable \mwis~application had two parts 
    (3x/day EMA surveys, and the corresponding visualization dashboard),
both administered via the HTP's standard website structure. 
Users  received text-message 
and email reminders to complete each survey. 

\inlineHeading{Impact Evaluation (\S\ref{Subsec:Results:UtilityStudy}, 4 months):}
Our clinical collaborators designed and administered a clinical feasibility study based on a three-week \mwis\ deployment with 44 youth with chronic pain, recruited through the clinicians' pain clinic. The A-B crossover quasi-experiment study design featured one week EMA-only and one week EMA+Visualization (order counterbalanced), with an intervening 1-week washout break. 
All patients in the EMA+Visualization condition saw just the current week's data, even if they received the EMA+Visualization condition last.
Separate publications describe this study's clinical protocol~\cite{boerner_data_2022} and clinical results~\cite{boerner_making_2023}. 

In this paper, we report on an N=11 impact evaluation that we conducted with a subset of the clinical study participants, 11 youth with chronic pain, interviewed at the end of the three-week trial period. 
As with the previous design validation, our methods consisted of semi-structured interviews and a questionnaire. In addition to a continued focus on overall understandability, we sought to understand the impact of \mwis\ during the week the participants saw the visualizations: which aspects were motivating or led to reduced use, and whether participants gained insights from the application regarding their pain and health. Youth were compensated CAD\$20 for their participation, in addition to the clinical study compensation of CAD\$40.

\inlineHeading{Reflections 
(\S\ref{Sec:DeploymentReflect}--\ref{Sec:DesignReflect}, 5 months at reduced intensity):} 
The clinical study concluded with sub-optimal and inconclusive results, largely due to significant technical issues with the HTP platform. 
In \textbf{Deployment Reflections (\S\ref{Sec:DeploymentReflect})}, we highlight the challenges we faced, the possible barriers that led to those challenges, and suggest mitigations that could be applied to overcome such challenges in future deployments with clinically vulnerable populations.
In \textbf{Design Reflections (\S\ref{Sec:DesignReflect})}, we consider the evolution and final version of the design to articulate design strategies that capture priorities and important tradeoffs we had to make.

\inlineHeading{Final Redesign (\S\ref{Sec:VizDesign} \& \S\ref{Sec:DesignReflect},  1 month):}
To continue this work, our clinician team needed to conduct a new clinical study with a new deployment partner. 
To create a handoff specification, we incorporated insights from the impact evaluation and our reflections on the 
design process (\S\ref{Sec:DesignReflect})
into a final design iteration (\protoname{Handoff}), creating mockups in Adobe Illustrator
in the same style as \mwis\ that show real patient data gathered in the previous studies (also covered in \S\ref{Sec:VizDesign}). 


\section{Requirements and Abstractions}
	\label{Sec:Abstractions}		

We describe a set of data and task abstractions that we iteratively developed based on requirements elicited from clinical experts and patient representatives (\S\ref{Sec:Process}).

\subsection{Data Abstraction}
	\label{Subsec:Abstractions:Data}	

Our dataset consisted of 3x-a-day responses to the EMA surveys (\SuppEMASurvey). 
We classified the EMA items into six \textbf{Facets} representing aspects of daily lived experience, detailed below alongside the EMA \textit{response options, in italics}.
%
Our designs incorporated only quantitative sequential (QS), quantitative cyclic (QC), ordinal sequential (OS), ordinal diverging (OD), and categorical (C) attributes, excluding free-form text responses. 

\inlineHeading{\textsc{Sleep}:} 
Clock times for going to sleep and waking up (QC); sleep quality (OS: \textit{poor, okay, good, great}), only for the first EMA of the day in the morning. 

\inlineHeading{\textsc{Symptoms}:} 
Type (C: \textit{stomach ache, headache, low back pain, dizziness, limb pain, fast heartbeat, nausea, body weakness, other}), intensity and level of worry about physical symptoms (QS: 0-10), and whether they took medication (C: binary). 

\inlineHeading{\textsc{Emotions}:} 
Types (\textit{worried, angry, happy, sad}; QS: 0-10). 

\inlineHeading{\textsc{Worries}:} 
The most worried about thing (C: \textit{family, friends, strangers, school, sports, health}),  level of worry, level of certainty that it would occur, and estimate of how bad it would be (QS: 0-10). In afternoon and evening EMAs: whether it did happen \& whether they attempted to avoid it (C: binary), and how bad it actually was (QS: 0-10). 

\inlineHeading{\textsc{School}:} In afternoon EMA, 
whether they attended school (C: binary) and if missed, the reason (C: \textit{weekend, holiday, vacation, pain, sick, medical appointment, home-schooled, online}). 

\inlineHeading{\textsc{Peers}:} 
Level of worry about peer interactions (QS: 0-10), their quality (OD: 5-level Likert), and occurence (C:binary).


\subsection{Requirements} 
	\label{Subsec:Abstractions:Requirements}		

We characterized and framed requirements iteratively, beginning with a thematic analysis of the 9 stakeholder interviews (\S\ref{Sec:Process}, stage 1),
followed by multiple rounds of feedback and evaluation all the way through Final Redesign.
We ultimately identified six high-level \goals\ for enabling youth to use the \mwis~application as a clinical intervention for reflection and management of their chronic pain, the first three drawn from these target users and their context, and the rest from the tasks being supported.


\inlineHeading{R1.\oldNumber{3} Motivate Teens to Engage:} 
We aspire to promote engagement from the target audience of teens at two levels: both to motivate their compliance with the data-gathering EMA protocol itself, and
to promote reflection on the meaning of that data, such as 
noticing relations between facets, or dissonance between how they recall their lived experience days or even hours later \vs how they logged it in the moment.
We posit that this reflection will lead to curiosity and micro-experimentation -- trying different daily choices to see if they yield better pain management or control; and ultimately inspire useful behaviour change.

\inlineHeading{R2.\oldNumber{5} Run on Mobile:} 
The application must run on a mobile phone, for availability during daily activities. At the device level, this dictates physical screen space, pixel resolutions, and computational power. Moreover, the mobile usage context frequently involves a multitasking or otherwise distracted user with a short attention span.  

\inlineHeading{R3.\oldNumber{4} Provide Information Quickly:}  
\mwis~should allow users to extract meaningful insights quickly, even within a single glance. It also must be immediately self-explanatory, since teens are unlikely to spend time reading instructions to learn how to interpret unfamiliar visual representations.

\inlineHeading{R4.\oldNumber{1} Distinguish Multiple Facets:} 
The survey data covers many different aspects of patients' lives, for a holistic scope that is deliberately broad. \mwis~should support focus and investigation on particular aspects of interest.

\inlineHeading{R5. \oldNumber{2} Support Cross-Facet Comparison:} 
Users should be able to reflect on how the different aspects of their life connect with and affect each other. This capability is a fundamental design target, although possibly in tension with other requirements 
	(R2, R3).  

\inlineHeading{R6. Handle Shallow Data:}
Although this dataset has many and varied attributes, each working snapshots is tiny 
by the usual standards of visualization, with just 21 timepoints for one week of data (3 EMAs/day). 
Moreover, cumulative collection meant fewer timepoints available early in the week, compounded by any lapses in survey-taking compliance. Enabling reflection and insight for such ``small data'' entailed interestingly different challenges from the common visualization research goal of scalability.

\subsection{Task Abstraction} 
	\label{Subsec:Abstractions:Tasks}	

Our opening thematic analysis (\S\ref{Sec:Process}) produced an initial \task~list with 
59
items.  
We grouped these within each of our six facets, or identified them as cross-facet tasks (see \SuppTaskAbs\ for the full \facet-categorized list.). 
It was not possible to fully support this comprehensive list within our scope and patient attentional constraints, so our clinical collaborators identified several factors as a framework to weigh relative overall importance of the candidate tasks.

 
\inlineHeading{User type:}
We deemed some tasks in our initial list likely to be most relevant for \textit{Clinicians}, while others focused on the interests of~\textit{Patients}, and some heavily overlapped. 
For instance, \textit{Correlate between type of Symptom and Emotions} would be more relevant for clinicians, but \textit{Identify trends in Sleep pattern} of greater interest to patients. 

\inlineHeading{Task type:} 
We classified tasks as \textit{Reflective} or \textit{Non-Reflective}, based on involving deeper engagement than simple lookup. 

\inlineHeading{Task clinical priority:}
Clinicians designated each candidate as \textit{High/Medium/Low} priority based on where they felt youth self-reflection would be most valuable in terms of the intended intervention.
 
\vspace{0.03in}
\noindent 
To keep the project within a realistic scope, we first eliminated all low-priority tasks; then retained \textit{Patient}-related tasks and eliminated most purely \textit{Clinician}-related ones, keeping some with substantial overlap to provide visualizations that catered to both groups, and assigning high weight to \textit{Reflective} tasks. 
Finally, we grouped these focus tasks into three high-level purposes. We list these abstracted tasks along with pointers to \S\ref{Sec:VizDesign} data descriptions, and mapped to data facets with the \mwis\ dashboard's color-codes (note that Emotions (5.3) is multicolor): 



\begin{center}
    \centering  \footnotesize
    \renewcommand{\arraystretch}{1.3}
     \label{tab:task_abstraction}

\begin{tabular}{p{.75\columnwidth} p{.15\columnwidth}}


\toprule

\multicolumn{2}{l}{\textbf{T1. Discover extremes.}}\\
T1.1 Periods of high-intensity pain & \styleSymptoms{5.2} \\
T1.2 Which symptoms are most frequent & \styleSymptoms{5.2}\\
T1.3 Which symptoms are most physically intense & \styleSymptoms{5.2}\\
T1.4 What worries the patient the most & \styleWorry{5.4}\\

\midrule
\multicolumn{2}{l}{\textbf{T2. Investigate trends over time.}}\\
T2.1 Sleep pattern & \styleSleep{5.1}\\
T2.2 Symptom occurrence & \styleSymptoms{5.2}\\
T2.3 Symptom intensity & \styleSymptoms{5.2}\\
T2.4 How worried they are & \styleWorry{5.4}\\
T2.5 If things they are worried about actually happen & \styleWorry{5.4}\\
T2.6 How bad things are actually versus expected & \styleWorry{5.4}\\
T2.7 Peer interaction pattern & \stylePeer{5.5}\\

\midrule
\multicolumn{2}{l}{\textbf{T3. Compare attributes.}}\\
\multicolumn{2}{l}{\textit{Investigate potential correlation within (W) and across (A) facets:}}\\
T3.1 How bad things are \textit{with} were they avoided (W) & \styleWorry{5.4}\\
T3.2 Medications taken \textit{for which} symptoms (W) & \styleSymptoms{5.2}\\
T3.3 Symptom intensity \textit{with} sleep (A) & \styleSleep{5.1}, \styleSymptoms{5.2}\\
T3.4 Symptom intensity \textit{with} emotions (A) & \styleSymptoms{5.2}, 5.3\\
T3.5 Symptom intensity \textit{with} peer interactions (A) & \styleSymptoms{5.2}, \stylePeer{5.5}\\
T3.6 Symptom intensity \textit{with} medication usage (W) & \styleSymptoms{5.2}\\
T3.7 How bad things actually are \textit{with} sleep (A) & \styleSleep{5.1}, \styleWorry{5.4}\\
T3.8 Worry about health \textit{with} symptom intensity (A) & \styleSymptoms{5.2}, \styleWorry{5.4}\\
T3.9 Symptom occurrence \textit{with} school attendance (A) & \styleSymptoms{5.2}, \stylePeer{5.5}\\

\bottomrule

\end{tabular}
\label{table:tasks}
\end{center}
\normalsize


\section{Visualization Design}
	\label{Sec:VizDesign}		


\begin{figure}[bt]
        \includegraphics[width=.5\textwidth]{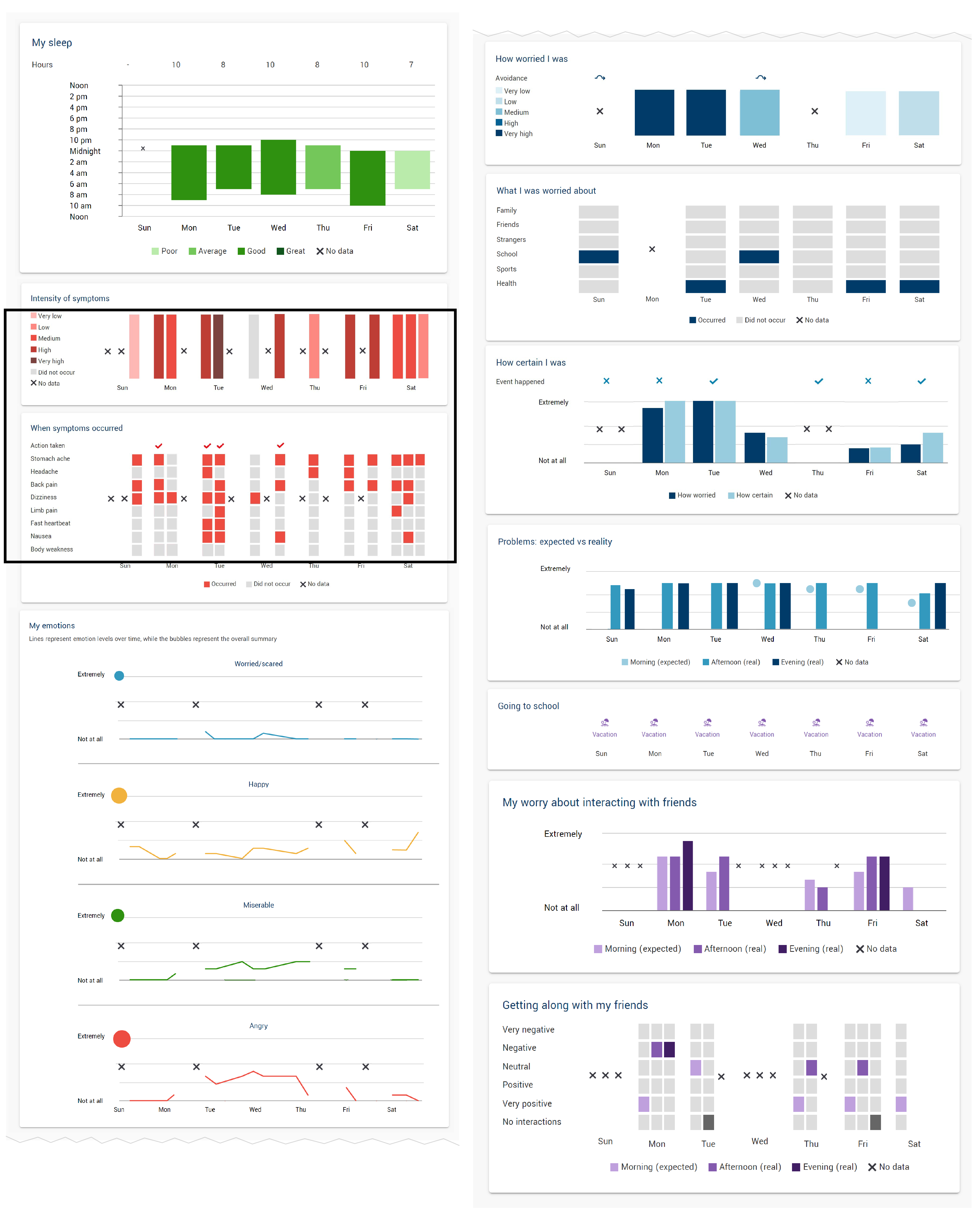}
    \caption{Visualization dashboard for deployed version (\texttt{Deploy}). Charts are vertically aligned, ordered and colour-coded by facet.  The full scrollable dashboard has a very tall (1:5) aspect ratio; hence, the figure is split into two columns. 
    This example combines data from multiple real participants, showing almost a full week of data. The black outlined frame 
    is an annotation indicating the screen extent of a smartphone in landscape
    orientation.
    }
    \label{fig:final_viz_dashboard}
\end{figure}

    \begin{figure}[t]
        \includegraphics[width=.5\textwidth]{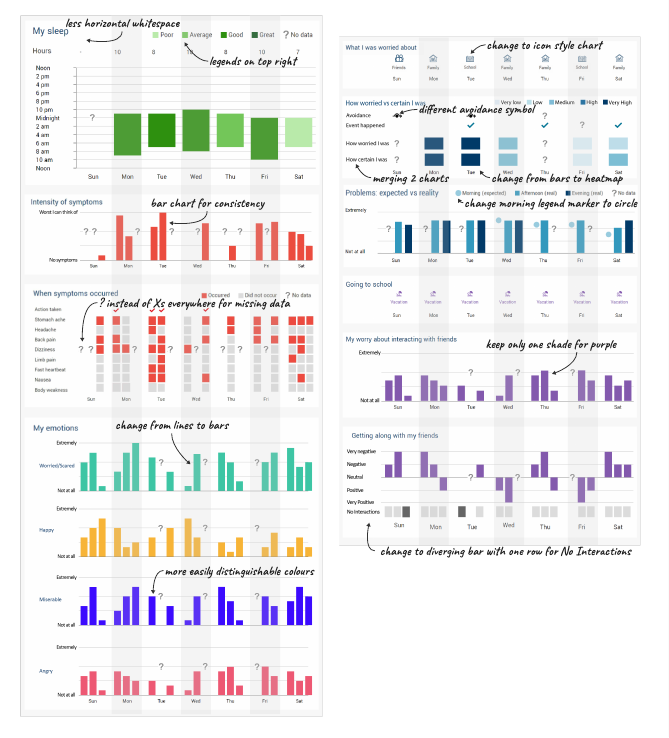}
        \caption{The final redesigned (\protoname{HandOff}) dashboard prototype, with changes from the \protoname{Deploy} version annotated. This example shows real participant data collaged into a mockup using Adobe Illustrator.}
        \label{fig:handoffPrototype}    
    \end{figure}


The \mwis~visualization dashboard is a suite of vertically aligned charts, grouped and coloured by \facet. 
The facet order was chosen to support the highest-priority comparison tasks (T3); \textsc{Sleep} is at the top, followed by \textsc{Symptoms} and \textsc{Emotions}, then \textsc{Worry}, \textsc{Peers}, and \textsc{School}. 
All charts feature time as the horizontal axis, with the entire week's worth of data visible at once, to support investigation of trends over time (T2). 
The only user interaction is vertical scrolling. 
At least two charts can be vertically juxtaposed in any orientation of the mobile device (portrait or landscape).
Missing data is indicated with a question mark symbol ('?').

\begin{figure*}
    \centering
    \includegraphics[width=\textwidth]{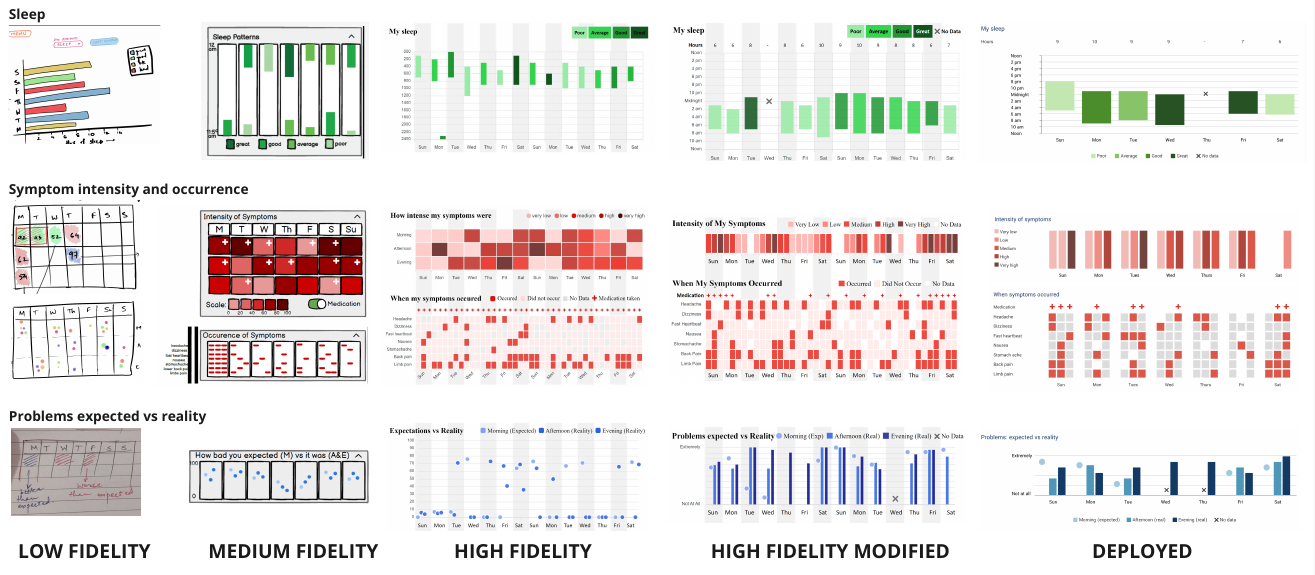}
    \caption{Evolution of three visualization \kmE{chart} designs, across \protoname{LoFi} (sketched), \protoname{MedFi} (Balsamiq), \protoname{HiFi} (React.js), \protoname{HiFiMod} (also React, after the Usability Pilot Study), and \protoname{Deploy} (after the Preliminary Utility Study) versions. 
    Rows show selected charts from the (a) \textsc{Sleep}, (b) \textsc{Symptoms} and (c) \textsc{Worries} facets.
    We initially prototyped visualizations for two-week EMA periods, but the deployed version shows only one week of data to support the finalized format of the clinical study.
    }
    \label{fig:evolution}    
\end{figure*}

Figures~\ref{fig:final_viz_dashboard}-\ref{fig:handoffPrototype} show the deployed  (\protoname{Deploy}) and final design (\protoname{Handoff}) versions. 
Figure~\ref{fig:evolution} shows the evolution for three facets across five prototype versions: \protoname{LoFi}, \protoname{MedFi}, \protoname{HiFi}, \protoname{HiFiMod}, and \protoname{Deploy}. 
\SuppVizChanges~details the evolution of all charts from  \protoname{HiFiMod} to  \protoname{Deploy} versions.

We now discuss the deployed and final designs of charts for each facet, the tasks they support, and their evolution across design stages. Most of our design choices support many of the cross-cutting requirements, so we will not enumerate these below. 

\subsection{\textsc{Sleep} Facet \styleSleep{(Green)}}
	\label{Subsec:VizDesign:Sleep}			

The final design of the single \textsc{Sleep} chart (\chartname{My sleep}) maps three key attributes of a single night's sleep naturally onto a simple base bar mark: sleep length as bar length, its start and end times as location on a daily calendar, and sleep quality by sequential discretized colour saturation. There is one answer each day, and the base hue is green. 

\inlineHeading{Tasks:} The \chartname{My sleep} chart supports the investigation of sleep pattern trends over time (T2.1). Its top placement supports the cross-facet comparison between sleep and symptoms, which was deemed most crucial (T3.3).
    
\inlineHeading{Evolution:} 
\chartname{My sleep}'s journey is featured in Figure~\ref{fig:evolution}(a).
We began (\protoname{LoFi}) with a standard barchart showing only durations, but the sleep/wake timing information was crucial, and integrating it from an adjacent or overlaid graphic added cognitive load.
The next (\protoname{MedFi}) version incorporated timing into the bar itself, with a conventional calendar: vertical time axis starting at midnight, also aligning to the other charts' temporal divisions.    
However, the confusion caused by marks split across midnight violated the glanceability requirement (R3), and also substantially increased perceptual difficulty of total-length judgements. 

Bar continuity thus took precedence in the next version (\protoname{HiFi}) and all later ones, where we chose a different calendar alignment. 
Sleep schedules can vary dramatically for youth with chronic pain, but our interviews suggested a Noon calendar boundary would minimize sleep bar splitting. 
We therefore reframed our interpretation of timing consistency: even when sleep occurs partly in the previous calendar day, it impacts the current calendar day and thus should be aligned with other current-day data. 

In the  \protoname{Deploy} version, bar mark progression quickly conveys the overall pattern of sleep schedule and quality. Sleep length is explicitly annotated through text labels at the top of each column, in response to user feedback after the usability pilot (\protoname{HiFiMod}); actual hours of sleep was the only absolute number of patient interest. 

Finally, in this chart and all others, we changed from a distractingly salient black 'X' glyph indicator of missing data in \protoname{Deploy} to a  more subtle grey question mark ('?') in \protoname{HandOff}. We also moved legends from below charts to the upper right corner for better information density.

\subsection{\textsc{Symptoms} Facet \styleSymptoms{(Red)}}
	\label{Subsec:VizDesign:PhysicalSymptoms}		

There are two physical \textsc{Symptoms} charts, both using a base hue of red. 
The upper \chartname{Intensity of symptoms} is a barchart showing magnitudes with vertical position. 
The lower \chartname{When symptoms occurred} chart uses a binary heatmap: 
presence (red) or absence (grey) of each of the 7 surveyed symptoms, on a categorical vertical axis. 
Absence is shown as grey rather than white to clearly distinguish it from the no-data case (empty white column with a single '?' in the middle). 
In the bottom \chartname{Occurrence} chart, checkmarks at the top of the column indicate if medication was taken that day.

\inlineHeading{Tasks:} 
These charts cover several survey questions and many tasks: within-facet discovery (T1.1-T1.3), progression over time (T2.2-T2.3), and within-facet comparison of medication usage to symptom type (T3.2) and intensity (T3.6). 

\inlineHeading{Evolution:} 
Figure~\ref{fig:evolution}(b) shows the evolution of both charts. The upper chart was a heatmap in all versions through \protoname{Deploy}, showing ordered Likert data with a sequential discrete saturation colour ramp, plus grey for no symptom reported. 
In the three earliest versions (Figure~\ref{fig:evolution}), the upper chart split a single column for each day into three stacked boxes for the three EMA time points, for the sake of high information density. The fourth prototype (\protoname{HiFiMod}) changed strategy to vertical rather than horizontal splits, to better align with the lower and other charts.
In the \protoname{HandOff} version, we made a major change for global consistency: we switched to a barchart so that all magnitudes gathered three times a day are shown in the same way. When no symptoms were reported, the column is simply blank (zero-height bars); when data is missing, the blank column contains a vertically centered question mark.

For these and all other charts, the first prototypes (\protoname{LoFi} and \protoname{MedFi}) outlined  each day in black for high-salience distinctions, whereas the \protoname{HiFi} and \protoname{HiFiMod} prototypes used alternating vertical grey / white stripes spanning the dashboard's full height to visually distinguish neighboring days less obtrusively and with less screenspace. 
With the fifth (\protoname{Deploy}) prototype, platform constraints prohibited the stripes; since the choice to show just one week's data during the study relaxed the very aggressive horizontal-space constraint, we switched to visually grouping via horizontal whitespace. 
However, it was considerably more difficult to track vertically across charts to compare values between them. In the final (\protoname{HandOff}) version, these constraints were removed, and we reinstated the grey stripes.
In the \protoname{HiFi} and \protoname{HiFiMod} versions, symptom-absence boxes were light pink, leading to potential confusion with the similar colour indicating \textit{very low} in the upper heatmap. We switched to grey absence boxes in the \protoname{Deploy} version, and ensured that there was sufficient luminance contrast between these and the background stripes in the final version.

\subsection{\textsc{Emotions} Facet (Multicolour)}
	\label{Subsec:VizDesign:Emotions}		

The \textsc{Emotions} facet design uses small multiples of the same chart,
since the identical question is asked for each of 4 emotions. 
The final version (\protoname{HandOff}) uses barcharts. 
Each emotion has a different colour (teal, gold, indigo, pink). Although these colors partially overlap with those of other facets, we felt that the benefit of maximal distinguishability within the facet outweighed concerns of potentially misleading associations across facets. 

\inlineHeading{Tasks:} These charts cover progression of worry over time (T2.4) and comparison of emotions to symptom intensity
(T3.4) through facet proximity. 

\inlineHeading{Evolution:} 
Although a barchart may initially seem an obvious choice, we did not start there in this facet either. 
All prototypes through \protoname{Deploy} used line charts to show emotion intensities, with bubbles representing the cumulative weekly total on the left side, superimposed in the same frame for compactness. Our original rationale was to emphasize trends over time of individual emotions and facilitate comparing between them, and to support design-time feedback that emotions feel 'continuous'. However, the EMA data is gathered as discrete samples at separate time points, so interpolation may not be appropriate.
Further, the unfortunate decrease of information density from 
{\protoname{HiFiMod} to \protoname{Deploy} (\SuppVizChanges) impeded the intended cross-facet comparison with symptom intensity above. 
Most importantly, the Impact Evaluation highlighted how significant missing 
data can dramatically impede the understandability of line charts: they become a field of cryptic segments. Figure~\ref{fig:final_viz_dashboard} shows one example, and the full set of participant dashboards in \SuppExamples~drives the point home. 

We thus switched from line to barcharts in our final redesign (\protoname{HandOff}), because in other charts we saw that they were more robust to missing data. 
Figure~\ref{fig:handoffPrototype}
shows the redesign as an Illustrator mockup using real data, demonstrating this much-improved resilience. The final version was also more compact, requiring $\sim$20\% less screen space.}
We shifted the colours slightly to be more distinguishable from the color palettes used in the other facets.

\subsection{\textsc{Worries} Facet \styleWorry{(Blue)}}
	\label{Subsec:VizDesign:Worries}		

The \textsc{Worries} facet contains three 
charts tracking different aspects of worry, with a base hue of blue. Their data is gathered less frequently than three times a day, in contrast to the other facets. 

In the final version, the first chart (\chartname{What I was worried about}) indicates the target of worries with icons and text labels. 
The second (\chartname{How worried}) compares their level of worry to their level of certainty about that worry, as heatmap tiles with a sequential colour coding; above them, reported avoidance behavior is indicated with a curved-arrow glyph, and whether the worried-about event did occur later that day is shown with a checkmark.





The last (\chartname{Problems: expected \vs reality}) uses a filled circle to show the morning expectation of how bad the main problem would be, and bars for the afternoon and evening reality of how bad it actually was. Our goal was to highlight potential cognitive distortions related to anxiety that may motivate unhelpful avoidance behaviour~\cite{Asmundson_Pediatric_2012}.


\inlineHeading{Tasks:} 
These charts support finding extreme values for worries (T1.4), and investigating several trends over time (T2.4-T2.6). They also support within-facet comparison of worries and avoidance actions (T3.1), and cross-facet comparison with sleep (T3.7) and symptoms (T3.8). These cross-facet tasks were deemed less important than others, so those facets were distant from this one.

\inlineHeading{Evolution:}
The top \chartname{How worried} chart in the \protoname{HandOff} version corresponds to a very different second chart in the \protoname{Deploy} version: a binary heatmap, similar in spirit to the \chartname{When symptoms occurred} chart, to support multiple simultaneous answers. The clinical team changed the question format to single-answer for the \protoname{HandOff} specification (and specified the need to see the accompanying free-form text as well). Thus, this chart could be more compact, echoing the \textsc{School} one described below.

The \protoname{HandOff} second chart merges two charts (first and third) in the \protoname{Deploy} prototype. We realized that the \protoname{Deploy} \chartname{How certain} barchart was globally inconsistent by using horizontal bar position to demarcate different questions rather than the same question asked at different times of day, as in all other charts. 
Moreover, the \chartname{How worried} data was redundantly shown in both charts. The combined redesign uses vertical position to distinguish between questions, for better global consistency of position coding; the vertical column of saturation-coded coloured tiles allows easy comparison between the worried and certain magnitudes generated each day. The avoidance glyph also changed from a carefully designed arrow curving around a circular obstacle, to evoke the idea of avoidance in the \protoname{HiFiMod} version, which could not be instantiated because of HTP technical constraints in the \protoname{Deploy} version (see \SuppVizChanges). We restored that desired level of visual polish in the \protoname{HandOff} version (Figure~\ref{fig:handoffPrototype}). 




\chartname{Problems: expected vs reality} evolved to emphasize asymmetries (Figure~\ref{fig:evolution}). Initially, all three values (morning expectation, afternoon reality, and evening reality) were shown as filled circles. We changed the mark types to distinguish the latter two, to emphasize the difference between estimation and actuality as time progresses. 


\subsection{\textsc{School} and \textsc{Peers} Facets \stylePeer{(Purple)}} 
	\label{Subsec:VizDesign:School}	
	\label{Subsec:VizDesign:PeerInteractions}	

\textsc{School} uses a single \chartname{Going to school} chart
combining icons and text labels to show attendance or various causes of absence (a once-daily question), with the usual day-by-day structure. 
It shares a hue of uniform purple with the \textsc{Peers} charts below it.
The first chart, \chartname{My worry about interacting with friends}, follows the first \textsc{Symptoms} chart by showing the intensity of worry about peer interactions with three grouped bars per day.
The second, \chartname{Getting along with my friends}, shows the actual reported quality of peer interactions over time.  In \protoname{HandOff}, the main data is shown with a diverging barchart. On the bottom row, the exception case of \textit{no interactions occurred} is shown as a binary heatmap with dark and light grey tiles.

\inlineHeading{Tasks:} These charts support the comparison of symptom occurrence and school attendance (T3.9), and the progression over time of peer interaction patterns (T2.7). Again, the symptom-school comparison task was deemed to be of lower importance than others; facet distance incurs some cognitive load.

\inlineHeading{Evolution:}
In the \protoname{Deploy} and earlier versions, the \chartname{Worry} and \chartname{Getting along} charts both used color saturation redundantly with horizontal position within daily bar groups to show time of day. In \protoname{HandOff}, we improved global consistency by dropping  to a single shade of purple, reserving saturation changes to show either magnitude or the heterogeneity of different questions at different times.

The \chartname{Getting along} chart changed substantially from the
\protoname{Deploy} version, where a heatmap showed magnitudes with vertical position, from negative (top) to positive (bottom), and time of day redundantly encoded with color and horizontal position.
The \protoname{MedFi} version of this chart had the opposite vertical ordering, with positive at the top and negative at the bottom. After the usability pilot, in version \protoname{HiFiMod} we changed valence for better consistency; the Worry chart in the \textsc{Emotions} facet has the positive outcome at the bottom. 
The original motivation for using a 2D tile matrix for \chartname{Getting along} was consistency with the \textsc{Symptoms} chart type. However, we identified a deeper, semantic inconsistency. For \protoname{HandOff}, we changed the encoding to show magnitudes with position through a barchart, reserving changes of vertical position in tile maps to distinguish categories. Although this choice introduced a new chart type -- diverging bars -- we felt semantic consistency was more important than minimizing chart types.

Finally, the intended vertical alignment of charts in this facet was not fully preserved by HTP in the \protoname{Deploy} version; this problem was corrected for \protoname{HandOff}.

\section{Formal Evaluation} 
	\label{Sec:StudyResults}	

We conducted formal studies at two stages (\S\ref{Sec:Process}, Figure~\ref{fig:process}): first to directly evaluate the visualization design, then to assess the utility impact of their in-situ use
(and the clinical impact, as reported elsewhere~\cite{boerner_data_2022,boerner_making_2023}).
These studies followed extensive formative consultation and feedback with domain experts and patient representatives.


\subsection{Design Evaluation Studies: High-Fidelity Prototype}
	\label{Subsec:Results:UtilPrelimStudy}	


We explicitly evaluated the two hi-fi (React.js) prototypes in the Design Evaluation stage, in two studies with intervening modifications to the prototypes (Figure~\ref{fig:process}):
the Usability Pilot with graduate students as proxy users (N=6), and the Preliminary Utility Study with the target population (N=10; 7F, 3M, aged 12-18 years). 
Both were based on 1-hour semi-structured interviews combining specific but open-ended questions, yielding qualitative data with a questionnaire for quantitative triangulation. 
In the latter, participants assessed each chart for Understandability, Utility, Interest, and Aesthetic appeal on 5-point diverging Likert scales, as well as Understandability of three pair-wise comparisons of charts drawn from different multi-chart facets (\textsc{Symptoms}, \textsc{Worry}, and \textsc{Peers}).
We analyzed qualitative interview data through affinity diagramming. 

All interview questions and the visualization screenshots that accompanied them are provided in \SuppDesignEval. 
Important design revisions arising from the Usability Pilot are discussed in \S\ref{Sec:VizDesign}; full changes are documented in \SuppVizChanges.

We focus here on the results of the Preliminary Utility Study, as it addressed the target population of youth with chronic pain. 
We did not look for participants' insights with respect to their own personal data since they were shown proxy rather than their own personal data in both studies.

The most germane feedback here pertained to Understandability: liked \vs confusing aspects of the visualization design.
For quantitative results, we highlight notable features (full plots in \SuppQuant). 
All \textbf{chart all-metric averages} were $>=$3.5 (chart-specific ratings $>$3.0),  
led by \chartname{Sleep} (4.5), \chartname{Symptom intensity} (4.2) and \chartname{Worries intensity} (4.1); lowest was \chartname{Quality of Peer Interactions} (3.5). 
Of specific \textbf{metrics}, Understandability did best across all charts (average 4.1; other charts averaging 3.8-3.9). It was  highest for \chartname{Sleep} (4.6), \chartname{School} (4.7) and lowest (3.4-3.5) for \chartname{Symptom Occurrence}, \chartname{Expectations \vs Reality} and \chartname{Quality of Peer Interactions}.
\textbf{Pairwise chart comparisons} were all weakly positive for Understandability ($\sim$3.5).


The qualitative data aligned with and contextualized the quantitative results.  Most charts were easily understandable and straightforward, providing evidence that we were at a reasonable stopping point in terms of the design evolution. 
An isolated point of confusion was the bubbles superimposed on the \chartname{My emotions} small-multiple line charts; once explained, participants felt it would be acceptable, and we  addressed this concern in the \protoname{Handoff} redesign.

Participants also provided considerable feedback on what kinds of data they found useful to track, including both requests for further data collection and disinterest in certain categories of data in the current EMA survey. Most notably, there were substantial individual differences in perception of data relevance. 3/10 termed \textsc{Worries} or \textsc{Peers} irrelevant, others noted the utility of \textsc{Sleep} (3/10) and \textsc{Symptoms} (3/10), and many asked for further symptom types or daytime naps to be tracked.
While useful for our clinical collaborators, such feedback on the scope of the EMA survey itself is orthogonal to the design considerations that we focus on in this paper. 
We do note (\eg for future customization-oriented iteration) the need for nuance: the primary intervention goal is precisely to help users see correlations they were not previously aware of.

\subsection{Utility Study: Deployed Prototype}
	\label{Subsec:Results:UtilityStudy} 		

The primary Utility Study was piggybacked onto the clinical study~\cite{boerner_data_2022,boerner_making_2023} (N=44) in the Impact Evaluation stage, where 11 participants (11F, aged 12-18 years) were interviewed at the end of their three-week trial period. The Utility Study aimed to assess overall visualization  impact and utility, asking participants about their experience with the representations of their own data that they saw 
during the clinical study.  

Our participants were all female. Our clinical collaborators confirmed a large gender imbalance in the population seen at their pain clinic, which is also known to occur in the general population of youth with chronic pain~\cite{king_epidemiology_2011}.

Methods followed \S\ref{Subsec:Results:UtilPrelimStudy}: semi-structured interviews with a questionnaire. 
The questionnaire differed only by some metrics, replacing Interest with Insightfulness and Accuracy. 
New interview questions asked whether aspects of the visualizations were motivating or discouraging, the relatability and authenticity of the visualizations in accurately reflecting participants' lived experience, and if \mwis~helped participants think of ways to manage pain or reflect on pain management strategies. We again analyzed the qualitative interview data with affinity diagramming. See \SuppImpactEval\ for semi-structured interview questions and the full questionnaire, including visualization screenshots. 

To summarize quantitative results: some metrics performed similarly to the previous study (full plots in \SuppQuant), as did ratings patterns for individual charts. Altogether they imply reasonable {usability} of the visualizations themselves. Modifications made for the deployed prototype did not seem to affect Understandability (averaging 4.1 across all charts in both); further, between-chart comparisons rose (3.8 compared to 3.5).  Aesthetics also held steady (3.8), and the new Accuracy metric averaged 3.6.

However, we saw impact of the technical and data quality problems with the deployed application (see \S\ref{Sec:DeploymentReflect}).
For example, Utility dropped from 3.8 to 3.4, and
the new Insights (2.8) metric was very low compared to the first study's Interest (3.9).
These changes might reflect identified technical issues (\eg delayed access to the visualizations), but could also reflect problems with the design.

    %

Qualitative feedback confirmed substantial individual differences between what data and visualization facets participants found relevant. 
Beyond this, several themes provided preliminary but encouraging evidence of utility for addressing the intended requirements and tasks. 

    \inlineHeading{Motivation for behaviour change:} 
    3/11 participants described how looking at the visualizations prompted them to try strategies to ``improve'' the visualizations by improving 
    their health (\eg by going to bed earlier to achieve darker and ``better'' sleep charts; or trying to minimize missing-data markers). One 
    reported trying new strategies to address their symptoms, \eg trying something different when they saw medications had not helped.
    
    \inlineHeading{Continued use:} 6/11 mentioned they liked using \mwis\ on a weekly basis for reflection;  9 said they would continue using the application for a longer period. 
    
    \inlineHeading{Accurate reflection of their lived experience:} 
    7 participants explained how the visualizations reflected their lived experiences at the time, and noted the importance of this accuracy for utility in future reflection.
        
    \inlineHeading{Motivation for compliance, in retrospect:} 
    7/11 participants missed seeing the dashboard during the study due to technical issues (\S\ref{Sec:DeploymentReflect}) or busy schedules, a ratio consistent with the larger clinical study~\cite{boerner_making_2023}. After seeing them during the interview, they speculated that the visualized data could have motivated better survey compliance. 
    
    \inlineHeading{Visualization utility:} 
    4/11  participants 
    said the visualizations were helpful in understanding their pain. 3 found patterns but were unsure how to interpret them, suggesting that some clinical support for visualization interpretation and discussion could be useful. 
    3 did not see any patterns, but
    we do not know (\S\ref{Sec:DeploymentReflect}) if these had seen or missed the visualization dashboard -- highlighting the need for strong data forensics capability in future deployments.

\vspace{0.03in} 
The qualitative feedback indicated that we were not fully successful in enabling cross-comparison across facets (R2). 
6/11 participants were unable to cross-compare the visualizations across facets, while one did not think of it as relevant. 
Participants also pointed out additional tasks they would have liked to accomplish, \eg customizing visualization choices and being able to share the dashboard with their care team from within the platform.  
Finding new strategies to facilitate comparison and supporting more tasks are additional directions for future work.

\section{Deployment Reflections} 
	\label{Sec:DeploymentReflect}		

\begin{figure*}  
        \centering
        \includegraphics[width=\textwidth]{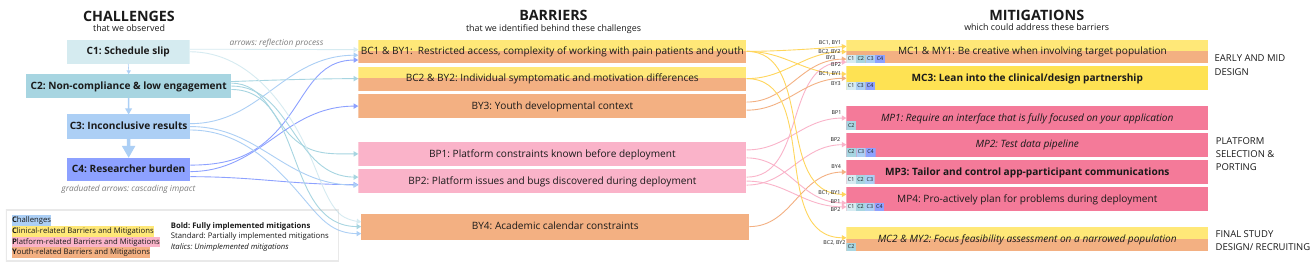}
        \caption{Challenges we faced, Barriers that might have caused them, and Mitigations that were or could have been used to overcome them, with their interconnections highlighted. Some Barriers and Mitigations relate to at least two  Clinical or Youth population or Platform factors. 
        }
        \label{fig:cbm}
    \end{figure*}

As we prepared to continue this work, we needed to guard against issues that impacted this first deployment. We clustered our challenges into four groups, connected these to the contextual and technical barriers that led to them, and for each barrier identified a set of mitigation strategies. 
We used the strategies proactively in selecting a new deployment partner and ongoing joint development; and have added implementation insights from this experience.
Figure~\ref{fig:cbm} maps these factors' influence on one another. We hope this framing can assist other researchers.

\subsection{Challenges}
	\label{Subsec:Deployment:Challenges}	

The four challenge clusters are cascading in their impact.

\inlineHeading{C1: Schedule slip.} 
Despite our team’s deep clinical experience with this population and our up-front awareness of the project's complexity, it took far longer to develop and deploy \mwis\ than anticipated. 
The final clinical/utility study was pushed to a time when teens were distracted and their routines disrupted, during end-of-school exams.
 
\inlineHeading{C2: Low compliance and engagement.} 
Survey completion rates were lower than expected. 
Post-study interviews indicated that many participants had minimally engaged with the visualizations, due to other distractions and engagements.

\inlineHeading{C3: Inconclusive results. 
The minimal data seen by some participants reduced opportunities for reflection sparked by} the visualizations.  
At a higher level, we obtained insufficient data on our approach's overall feasibility to fully assess its suitability as a health intervention. 

\inlineHeading{C4: Researcher burden:} 
The intersectional complexities of doing research with clinical and youth populations, and of deploying a technical system in the wild, caused 
substantial overhead for the team.  

\subsection{Barriers}  
	\label{Subsec:Deployment:Barriers}		


We identified eight 
barriers contributing to these challenges, further classified as Clinical (C), Youth (Y), and Platform (P). A barrier was often a factor in multiple challenges, as seen in Figure~\ref{fig:cbm}'s many arrows. 
We discuss barriers ordered by the chronological phase where they emerged in our design and deployment process.



\inlineHeading{BC1 \& BY1: Restricted access to and complexity of working with clinical patients, and youth.} 
With vulnerable individuals (here, clinically diagnosed with chronic pain \textit{and} underage), researchers face access barriers to getting their crucial feedback and incorporating it into a design process. 

\lowHeading{Clinical population (BC1)} 
    Time and effort requested of patients with chronic pain cannot impose an undue burden, 
    including anxiety arising from self-reflection. We needed to watch for adverse responses as the study proceeded and limit the number and length of participant interactions. 
    Meanwhile, clinical ethics approval is 
    extensive and inflexible compared to the behavioural ethics typically used in HCI research.  
    
\lowHeading{Youth (BY1)}
    Research with minors complicates and lengthens study administration, \eg consent is required from both patient and parent, through discussions that often need multiple interviews.

\inlineHeading{BC2 \& BY2: Individual differences.}
Our participants had high variability in symptom presentation and motivation.
    
\lowHeading{Symptom presentation (BC2)}
    Wide symptom variation in youth with chronic pain leads to extensive diversity in patient experience~\cite{cushing_tailoring_2019,Fillingim_Individual_2017}. Unable to encompass all experiential possibilities in our design, we focused on those identified by our clinicians as most common. 
    Unsurprisingly, post-study feedback showed that patients wanted to track and view other parameters, but we did not predict how disengaging this limit on personal relevance would be, even when participants were aware that they were giving feedback on a ``concept'' system.

\lowHeading{Motivation (BY2)}
    In early interviews, we found substantial variability in teens’ self-awareness, ability to reflect on personal patterns, and motivation to improve habits related to their health and well-being, resulting in wide-ranging motives and attitudes about \mwis. In deployment, participants whose needs diverged from \mwis's coverage were less engaged. Our sample was too small to assess preference prevalence or clusters, and catering to all needs was infeasible.

\inlineHeading{BY3: Youth developmental context.} 
    School, friends, managing and regulating emotions, and worries about academic or sports activities can be more relevant to their pain symptoms for adolescents than for adults~\cite{Rosenbloom_Developmental_2017}. Keeping the teen mindset central during design can be difficult for adult researchers, with their different current lived experiences. 


\inlineHeading{BP1 \& BP2: Platform constraints, issues \& bugs.}
The Impact Evaluation interviews indicated that technical deployment issues caused confusion and reduced compliance. 

\lowHeading{Discovered before deployment (BP1)} 
    Some inherent limitations became evident after the partnership sign-on with our HTP. We addressed some through redesign from our side and additional development by the HTP, but with substantial delay to the deployment timeline. Others, such as the HTP's requirement for weekend start, affected compliance. 
    
\lowHeading{Discovered during deployment (BP2)}
Some participants reported usability difficulties in finding the interface's visualization dashboard, and consequently did not interact with it for all or part of the EMA+visualization week. 
From a data perspective, during the clinical study period, we found that some EMA survey responses had ended up being incorrectly stored, leading to missing data even for participants trying to comply; the resulting frustration had contributed to later non-compliance. 

\inlineHeading{BY4: Academic calendar constraints.}
The schedule slip forced us to run the study at the end of the school year (June/July 2022) when potential participants were busy with exams. This timing made it harder to recruit and contributed to more delays. When teens did participate, survey completion and platform engagement were lower than they would have been at an earlier point, as identified via post-study 
interview statements. 
Participants who began after exams ended stated they were able to comply due to summer schedule flexibility.

\subsection{Mitigations}
	\label{Subsec:Deployment:Mitigations}		

We list identified mitigations in the same chronological order as for Barriers, indicating the best points to proactively implement each one.
As indicated, we incorporated some of these strategies  fully into our design and deployment process, others partially or on-the-fly late in the study; several are retrospective realizations.

\inlineHeading{MC1 \& MY1: Opportunistically triage involvement of intersectionally vulnerable users in design and evaluation 
\textit{[Used]}}.
User-centering is obviously crucial in designing for a vulnerable population; 
    \eg few adults can speak to a teen patient's perspective (C3). However, obstacles to user access are daunting (BY1), so a deeply engaged method (\eg participatory design) 
    may be infeasible (BC1).
    Instead, designers must adapt flexibly to users' lives, energy and available recruiting pathways.

    In our case, we minimized individual asks and conserved our limited sample with a rolling-evaluation strategy, recruiting youths with chronic pain for one-off feedback sessions ``whenever we can get you'' decoupled from specific milestones.
    This spread-out method nevertheless provided a first-hand sample of individual symptomatic and motivation differences (BC2, BY2), and many actionable design insights from a youth perspective (BY3). 
    It should usually be possible to prioritize scarce target-user time, addressing technical and general usability issues with proxy testers, domain experts (BP2) and any available heuristics.

\inlineHeading{MC3: Lean into a clinical / design partnership
\textit{[Used]}.} 
    HCI designers are accustomed to going directly to their users. 
    For clinical populations, however, clinician partners can be far more than stand-ins for tough-to-reach patients. 
    They provide a medically-informed perspective on their patients' concerns and needs (BY3), 
    steer projects through rigid clinical evaluation and ethical processes, and  provide access to  patient populations (BC1, BY1) 
    which would otherwise be off-limits. 
    They may also have access to funding sources attuned to lengthy, constrained studies. 


\inlineHeading{MP1: Require an interface fully focused on the application 
\textit{[Untried]}.} 
    The platform should allow a user experience free of irrelevant widgets and dead-end stubs. 
    Due to budget and timing, we accepted a pre-configured platform which came with distractors that caused misdirection and confusion (BP1), and ultimately contributed to non-compliance and low engagement (C2).

\inlineHeading{MP2: Test data pipeline 
\textit{[Untried]}.} 
    It is crucial to debug and pilot the deployed data pipeline \textit{separately} from interactive elements, and the research team must be engaged in this. 
    Even more proactively, a potential platform partner would provide a test application that exercises the full data pipeline before full commitment (BP2). 
    
    For post hoc analysis, the platform should provide fine-grained interaction data, \eg user screen views and action logs, so the team can identify technical issues and ground research in data. 
    Together, up-front verification and post-hoc analytic tools confirm data security and reduce researcher burden (C4) and inconclusive results due to bugs, technical issues and uncertainty about what users did or saw (C3).

\inlineHeading{MP3: Tailor and tightly control app-participant communications
\textit{[Partially Used]}.} 
    In general, users will use this type of app 
    only when they are effectively reminded to, especially over longer periods, and when app use is during daily life (BY4). The ideal contact modality may be demographic-specific (for teens, texts are usually more effective than email). T
    iming is critical, at two levels:  the respondent may have better and worse response windows, and the research objective will require some level of precision. 
    Thus, it is imperative that the platform gives researchers sufficient control over notification modality and timing (tailoring to individual schedules if needed), and that this functionality is on-time and bug-free in all conceivable situations.
    We achieved text-message reminders, but had limited and bug-prone control over their timing, undermining compliance (C2).
    We were able to push text notifications for the EMA surveys 3x/day, but in some cases they were delayed or not received, leading to lower compliance (C2).

\inlineHeading{MP4: Pro-actively plan for deployment problems
\textit{[Partially Used]}.} 
    Few early, feasibility-focused deployments fully avoid usability or performance issues that could jeopardize research findings. 
    Some defensive lines are possible.

    Pre-deployment, the team can prepare for user self-help  (accessible FAQs, guides, tooltips for understanding, navigating, troubleshooting); and/or access to ``live'' technical help from a team member. 
    We also suggest curating an onboarding experience for the first time that users access the app: interactive demos, tutorial videos or walkthroughs that familiarize users with the app workflow. Beyond guiding intended use, it could head off known usability issues that are impractical to address in a feasibility deployment. 
    By reducing confusion or helping to address platform limitations, this strategy can decrease non-compliance and researcher burden (C2, C4), in turn decreasing schedule delays and inconclusive results (C1, C3).

    Once deployed, fixes need to be implemented on-the-go, ideally with rolling recruitment so that participant experience can be improved as the study continues (if this does not jeopardize study condition consistency). 
    The platform partner must be able to respond quickly.

    For example, we sent out a guide to the platform and study process a day before the study began to the participants, to familiarize them. They were also encouraged to email the research team in case of any issues (some did), and our HTP enacted some fixes on-the-go through our rolling recruitment process, \eg reminders on where to view the visualization dashboard when we learnt participants were not able to find it. 

    Together, these strategies can reduce the impact of delays (BP1) and bugs (BP2) and are all the more valuable when each participant's involvement is hard-won (BC1, BY1). While such measures have an implementation cost, overall researcher burden (C4) may be reduced especially if planned for, while greatly lowering the risk of debacle. 


\inlineHeading{MC2 \& MY2: Focus feasibility assessment on a narrowed population.}
    Given variability in key factors (\eg symptom presentation, motivation) across a group, a common compromise is to narrow a small study's sample for a better view of feasibility, albeit at risk of missing larger patterns.
    
\noindent\textbf{\textit{Tighten symptom range (MC2) [Untried]:} }
    Chronic pain can present through a diverse range of symptoms, many relevant to few users (BC2); supporting the full range may undermine engagement (C2). 
    Tailoring app details to a narrower demographic (here, type of chronic pain diagnosis or primary pain site) could reduce response variability and allow assessment of core app functionality with users who find it personally relevant. 
    
\noindent\textbf{\textit{Narrow age range (MY2) [Untried]:}} 
    Post-deployment interviews underscored the very diverse needs, emotional and cognitive capacities, and experiences of teens across 12-18 years. Younger users found it complex, and were less willing to engage with the app (C2). We conjecture that design focused on a narrower age group could be more engaging.

\section{Design Reflections} 
    \label{Sec:DesignReflect}

We discuss successful strategies that we articulate as design principles, then design surprises, and finally some open questions.

\subsection{Successes, Captured as Design Principles}
\label{Sec:Subsec:DesignSuccesses}

We start by articulating several design principles (DP), many of which cross-cut all requirements and high-level tasks.
The results of our formal evaluation provide us with sufficient confidence to deem these successful.

\inlineHeading{DP1: Link Charts via Global Positional Alignment.}
Despite our data's heterogeneity, we were able to impose a consistent positional framework.
All charts share a horizontal time axis, ordered temporally and discretized at two levels: days of the week (coarse) and 3 times within each day (fine). 
The small number of time points, under 21 or 42, had the benefit of allowing sufficient information density so that the entire range could be visible simultaneously for per-attribute overviews.
This shared horizontal axis allowed all charts to be vertically aligned to support comparison~\cite{crisan_2022}, both within and across facets. 


\inlineHeading{DP2: 
Prioritize Key Tasks with Vertical Chart Ordering.}
A central challenge of the small mobile screen is that the full suite of charts cannot be seen simultaneously. We chose a vertical ordering for the charts to prioritize the most important tasks with vertical adjacency. Charts are vertically grouped together by facet because within-facet comparisons are the most crucial tasks. The initial vertical order across facets was carefully chosen according to clinician-specified priorities, and further tuned according to patient feedback from the design evaluation study. 
The resulting vertical ordering for \protoname{Deploy}, with 
\textsc{Symptoms} beneath \textsc{Sleep} and above \textsc{Emotions}, 
followed by \textsc{Worries} then \textsc{School/Peers}, was sufficiently robust that we made no changes for \protoname{HandOff}.

Figure~\ref{fig:final_viz_dashboard} highlights the inversion of aspect ratios between the visualization's whole and its parts: the full scrollable dashboard is $>$5$X$ taller than wide, while individual charts are 2--3$X$ wider than tall. 
Our design supported phone usage at both orientations.
\SuppExamples~documents the screen extent for each, typically with three charts visible simultaneously with portrait and at least two in landscape.

\inlineHeading{DP3: 
Distinguish Between Facets with Colour.}
The categorical colour coding to connect charts to their parent facets reinforced the grouping into facets by vertical position. This double encoding was particularly important because of the limited number of charts visible at once on the mobile screen, so absolute position could be difficult to detect. The four base hues were chosen to be distinguishable from each other and suitable for saturation-coded ramps against a white background. The exception was \textsc{Emotions}, with its identical small-multiple visual encodings, where colour was used to distinguish between those charts within the facet. 


\inlineHeading{DP4: 
Avoid Interaction Costs through Static, Scroll-Only Charts.}
Vertical scrolling was the only supported interaction,
thus avoiding the time and cognitive costs of interactive selection or more complex navigation~\cite{lam_2008}. 
This simplicity also avoided the learning costs of discovering interaction widget functionality.
This static approach aligned with previous study findings on visual encodings for mobiles which showed small multiples improved over animation for showing trends on small screens~\cite{brehmer_2020}. 

\inlineHeading{DP5: 
Indicate Missing Data with Appropriate Salience.}
We deliberately emphasize the places where data is missing with a glyph
against a white background. We rejected the common approaches of downplaying gaps or attempting to impute missing data, instead drawing attention to these gaps in hopes of motivating compliance in future EMAs; participant feedback indeed noted the motivating force of seeing these explicit marks. However, after seeing the extent of missing data in many participant dashboards, we moderated  salience level. 
We feared that the highly salient black 'X' could convey finality and even provoke anxiety, and hoped the more subtle grey '?' glyph in \protoname{HandOff} would pique curiosity and thus spur participants to better compliance when completing future surveys.

\subsection{Surprises}

When reflecting on design principles, we also identified several design surprises (DS) which we articulate for their nuances on conventional wisdom.  

\inlineHeading{DS1: The Complexities of Achieving Consistency.}
The design principle of aiming for consistency is surely unsurprising; most visualization designers would consider this idea obvious. The surprise lay in the difficulty of achieving consistency: we did not expect to discover major inconsistencies so late in the design cycle, after so many previous rounds of iteration and consideration.

At early project stages, we deliberately pursued some divergence in chart types to investigate whether chart type variety might promote engagement. Study feedback pointed instead to prioritizing consistency to avert confusion and maximize learnability. 

However, the legacy of the initial divergence continued to shape our thinking in ways we did not fully realize until much later.
Despite extensive scrutiny at five full stages of design from both the design team and the clinical team, and detailed analysis of feedback from four user studies, we did not realize several instances of global inconsistencies until the intensive reflection and writing at the end of the Impact Evaluation phase. Our experience aligns with and underscores the framing of \textit{writing as research}~\cite{sedlmair_2012}.

The primary shifts in the final \protoname{HandOff} design involved the position and colour channels. We standardized on encoding all magnitudes gathered three times a day exactly and only through the vertical position channel in barcharts, eliminating saturation coding within coloured-tile heatmaps and position coding with line charts for these cases. Conversely, we eliminated grouped barcharts for different magnitudes gathered only once a day because horizontal position did not exactly align with time of the response; instead we standardized on saturation coding within coloured-tile heatmaps for this case. We also eliminated colour changes that simply double-encoded the same attribute as position. Within a barchart, colour changes are now fully reserved to emphasize the asymmetry of different questions at different times. In the \textsc{Sleep} chart, colour changes encode a different attribute than the one shown with length.

The eventual convergence of our design to incorporate so many barcharts is somewhat surprising, in the face of calls for expressive layouts to promote engagement~\cite{brehmer_2017}  and moving beyond precision-driven visualization~\cite{bertini_2020}. The driving benefit of this chart-type convergence is a reduction of cognitive load and learning overhead.

%

In retrospect, two major factors combined to keep the design team from catching those global inconsistencies earlier: localized scrutiny and complex interrelationships between data attributes. Charts were carefully developed individually to visually encode the specific set of attributes they contained; there were frequent chart-level design critiques to examine each chart's utility, with substantial chart-level iteration to refine each of them. Charts were most frequently compared to those within the same facet to consider facet-level cohesion. We did consider global consistency issues all along: as we articulated the more global design principles in \S\ref{Sec:Subsec:DesignSuccesses} we also checked for consistency according to them. Although these principles did explicitly involve the channels of position and colour, they did not suffice to reveal the problems we eventually uncovered. 
Early on, the distinction between responses elicited symmetrically versus differentially across the three timepoints of each day appeared to be a somewhat peripheral detail; we gradually understood it to be a central semantic consideration. In the final \protoname{HandOff} version, the cases where different questions are asked as the day progresses are all visually emphasized and distinguished, to promote reflection and calibration.

\inlineHeading{DS2: The Ambiguity of Null Values.}
The treatment of null values could be considered a special case of consistency; we call it out, in particular, to illustrate how these seemingly similar data values all illustrate different phenomena depending on the task and data context.
The primary cause of this complexity is the number and heterogeneity of the data types viewed together. 
For example, in \protoname{Deploy}, \chartname{How certain I was}
used a \textit{blue} [\checkmark/X] to indicate whether a worried-about event did or did not happen, and blank for no response. In the other barcharts, a \textit{black} 'X' meant a skipped entry, and a blank (zero-height bar) meant a supplied '0' value. These cross-chart conflicts in use of the 'X' glyph and empty space arose because within the chart each choice seemed optimal, but globally became confusingly inconsistent.
This issue was exacerbated by the prevalence of zeros and misses in our real data.

Other ambiguities only became apparent after looking at real collected data, often related to sparsity. For example, on the \chartname{How Certain I was} chart, we realized that X's \vs checks became misleading when there were only a few data points. In reviewing real datasets post-hoc, we sometimes had to consult raw data logs to untangle ambiguities. Although participants did not complain, we took the opportunity to avoid these problems with the \protoname{Handoff} redesign.

\inlineHeading{DS3: The Brittleness of Line Charts.}
Line charts are a flexible and common chart type; we did not expect to encounter perceptual difficulties with their use. The real data that we collected late in the project, only at the Impact Evaluation study stage, exposed a number of problems relating to missing data that were not apparent in earlier stages. 

In some cases so much data was missing that the many disconnected segments were difficult to even recognize as a line chart; in less extreme cases, the chart type was clear but small differences were hard to notice across the gaps. 

The proportion of missing to available data was dramatically higher in the real data than in the test datasets we used during development, in part due to the platform issues barrier (BP1 \& BP2, \S\ref{Subsec:Deployment:Barriers}). Also, it showed data corresponding to the end of the collection period, but the study design intrinsically guarantees that there will be very little data available at the beginning. 


Late data access is a known hazard, explicitly identified 
as a pitfall to be avoided: 
\textit{no real data available (yet)} is suggested as a reason to decline a potential collaboration early on in a project's life cycle~\cite{sedlmair_2012}. Late data access was no surprise: it was understood from the project start, arising from the restricted access to the target population barriers (BY1 \& BC1) identified in \S\ref{Subsec:Deployment:Barriers}; we chose to pursue the project nevertheless. 


\inlineHeading{DS4: The Challenge of Shallow but Broad Data.}
Deep datasets, with many time points, are a well-studied scalability challenge in visualization research. Although the number of time points in our dataset is very modest, we still faced interesting design challenges. We characterize this visualization design problem as handling shallow (few time points) but broad (many heterogeneous components) data.

The 23 EMA response options, which we grouped into six \facets, constituted a broad set of heterogeneous data attributes, with many direct interrelationships and task-related interconnections between responses both within and across these facets. The heterogeneity precluded a uniform visual encoding: we had to approach chart design for each facet independently, identifying the central concept of interest for each one and iteratively constructing a visual representation that incorporated key attributes in a holistic, glanceable way. 
To support reflection and pattern-finding with small, recently self-reported data, we emphasized relative comparison rather than precise value identification. 

As discussed through the other design surprises, our extensive formative and summative evaluation process did provide useful feedback about the immediate legibility of our visualization designs, despite the technical obstacles barring some users from accessing and making sense of their own data 
(\S\ref{Sec:DeploymentReflect}). We reflect on the two extremes of chart understandability, high and low.

The \textsc{Sleep} chart worked well. We attribute this success at least partly to careful multi-attribute design and our very extensive iteration process.
This chart is unique yet easily understood, 
with some adherence to the basic form of day-linked bars featured in other charts plus modifications to the basic visual language to express multiple attributes at once.
However, ``sleep quality'' is inherently a graspable notion. Our teens could comprehend its attributes, were aware of its variability, and open to its being affected by other factors. Its concrete and quantifiable attributes are amenable to graphical interpretation.

The \textsc{Expectations \vs Reality} chart has none of these helpful traits. We labored on its representation (Figure~\ref{fig:evolution}), and yet it was poorly received in our user studies, particularly in perceived accuracy. 
However, it was considered crucial from an intervention standpoint, and scored highest of all our charts for \textit{potential utility}; teens evidently found it intriguing. Was the inaccuracy with the expectations measure, or the reality? Was the visualization just poorly executed (plausible, given improvements found for the \protoname{Handoff} version); or as an unfamiliar and amorphous concept, just tougher to portray? 
Much of our language and ideas come from adult clinicians, and may not fully reflect teen patient reality. There may be more relatable ways of expressing ``expectations'' and ''reality''.
More generally, the wording of other questions and even what factors may influence pain or wellness, and should be covered in the EMA, may be improveable.



\subsection{Open Questions}

We now reflect on questions that remain open. 

\inlineHeading{OQ1: Engagement and Personalization.}
    \label{SSubsec:DesReflect:Imp:Engage}
Providing motivation for teens to engage is an explicit requirement (R1) and a central concern. Assessing success is tricky because of the bidirectional dependencies between the data gathering through the EMA and the visualization interface to show that data. We are aiming for a positive cycle of effort and insight, where improvements in data-entry compliance are driven by interest in the data for what it tells users about themselves when they reflect on the visualizations. 
Our project attempts to prime this pump with visualization -- there is no insight without some effort, when the data is self-gathered. However, a negative spiral is also possible, where the visualization is unrewarding to use if the previously gathered data is too skimpy to provide insight, leading to further dropoff.
This confounding factor makes it difficult to draw definitive conclusions about the efficacy of the visualization design, but our findings help us frame further questions. 

We wanted to gauge how much the EMA structure aligned with the interests of the teen participants and consider what to do where those diverge. 
As we note with the Individual differences barriers (BC2 \& BY2), there was  high variability of both experiences and desires within this group. Our findings imply that even the mitigation strategy of focusing on a narrowed population (MC2 \& MY2) would only partially address the divergences.

The design team treated the EMA structure as unchangeable within the scope of this project and designed the visualization to support exactly that, but the study findings may lead the clinicians to refine it in future work.
Data entry requires effort up front for potential future benefit from insights;
some reward clearly needs to come soon and without perfect compliance.
Reducing the EMA to be more selective about the data collected was requested by some teens. 
Reducing the EMA burden could improve sustainability, and hence chances of users logging enough data to be interesting. We saw some evidence of ``bingers'' (high followed by zero compliance) \vs ``minimal completers'' who filed something at almost every cycle, if not complete. 

Another possible solution is personalization to increase perceived relevance, 
to control what questions are asked in the EMA or how answers are shown in the visualization.
However, the cost of patients only seeing what they are looking for is that they may miss valuable but unexpected insights.
For example, some teens did not report finding the \textsc{Worries} facet valuable, but the large amount of screen space devoted to it was endorsed by our clinicians. They did suspect that its importance was likely to be perceived as low but deemed it to actually have high importance in the lives of the targeted teens, thus providing an opportunity for constructive cross-comparisons. 

While we have focused here on teen motivation and insight, our patients' caregivers are also critical stakeholders.
Another level of motivation for compliance might be social in nature, \eg team-building and joint sense-making with their clinician to help a patient understand the data's value to their own health.

At the visualization level, the user might simply re-cluster or re-order charts to optimize comparison -- facilitating user-directed experimentation.
We originally planned for reordering, but dropped it due to our deployment platform's constraints; we would like to introduce that capability in a future iteration.

Behaviour modification is a long-term engagement objective; our studies only measured very short-term behaviour changes as a response to data exposure, providing some hints towards the visualizations' intrinsic and long-term value. Our principle of motivating compliance with visualizations that highlight rather than accommodate missing data (DP5) is a strategy explicitly aimed at direct behaviour influence, urging users to ``complete a streak’’. This feature did seem to motivate participants, at least in the short term, to provide more complete EMA data.


 
\inlineHeading{OQ2: Impact of Aesthetics.}
The design team bemoaned a steady drop in the teen-friendly aesthetic  between early sketches and the \protoname{Deploy} version (Figure~\ref{fig:evolution}), as our development platform become more constrained, rigid and business-toned; we feared the visual result was less fun and appealing. 
This shortfall showed up in aesthetic style, visual polish, and specific details such as icon design.
While we believe that our \textit{designs} improved with iteration, the real-world exigencies of changing platform constraints also brought degradation.
Although we were not able to definitively verify it from this project, we speculate on a relationship between aesthetic appeal and engagement, and see some indiciations of this in participant responses.


\inlineHeading{OQ3: Potential Generalizeability.}
    \label{SSubsec:DesReflect:Imp:Generalize}
Many of our design principles, surprises, and questions are relevant beyond the intersectionally vulnerable audience of teens with chronic pain: other audiences with potentially less stringent requirements may still benefit from our findings. 
For instance, teens may want to track aspects of their lives that are not focused on pain. Conversely, adults with chronic pain may want to understand condition triggers (R4) with minimal cognitive effort (R3, R5) and may also value the freedom of reflection on results in the moment (R2). More broadly, our findings may apply to other kinds of personal data sensemaking with shallow and broad self-generated data. Even more broadly, these ideas could apply to any kind of multi-facet comparative data task executed on-the-go. 



\section{Conclusions and Future Work}
	\label{Sec:Conclusion}

We have described the design and deployment of a visualization tool aimed at enabling youth with chronic pain in symptom self-management, to eventually serve as a treatment intervention. 
We developed design requirements, task and data abstractions, and full design and rationale for a multifaceted set of personal data visualizations, validated through extensive evaluation 
including a clinical deployment that demonstrated their potential. 
We provide a discussion on longitudinal clinical deployments of personal data visualizations, including a framework to structure our experiences in mitigating the challenges and barriers we met. We also reflected on our design experience, with a discussion of successful design principles, design surprises, and open questions. 

Our participants' feedback was overall enthusiastic and positive, with most highlighting the personal utility of such visualizations and saying they would use this system to manage their symptoms.  Visualizations that encode personal health data in engaging and youth-friendly formats clearly can seem relevant to teens in a clinically useful way.



\section*{Supplemental Materials Index}
Our supplemental materials are available on \textcolor{blue}{
\url{https://osf.io/af9p3/?view_only=c1ab0c23525b44e589eb7770f57f5a09}}
and include 
the early task abstraction (\SuppTaskAbs), changes for deployment (\SuppVizChanges), the \protoname{Deploy} (\SuppVizFinal) and \protoname{HandOff} (\SuppVizRedesign) interfaces, EMA Survey (\SuppEMASurvey), study materials  
(\SuppDesignEval, \SuppImpactEval), all participant dashboards (\SuppExamples), and quantitative study results (\SuppQuant).




\section*{Acknowledgments}
R\'{u}bia Guerra, Katra Farah, Haomiao Zhang, and Devarsh Bhonde contributed to the first \mwis\ design iteration. 
We thank Javed Gill and Jessica Luu, our many colleagues who offered advice and feedback, and above all the 54 teen patients, family members and advocates who offered specific feedback and participated in our studies at considerable personal effort.
This work was primarily funded by the BC Children's Hospital Research Institute and UBC's Designing for People NSERC-funded CREATE grant.
The reported studies were conducted under UBC's Behavioral Ethics Application H20-02965.

\bibliographystyle{IEEEtran}
\bibliography{IEEEabrv, mwis}









\vspace{-33pt}
\begin{IEEEbiography}[{\includegraphics[width=1in,height=1.25in,clip,keepaspectratio]{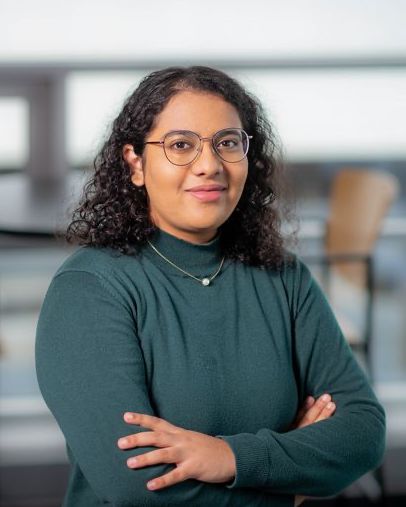}}]{Unma Desai}
is a UX Researcher and Designer. She received her M.Sc. in Computer Science from the University of British Columbia in 2022, and her Bachelors in Computer Engineering from the University of Mumbai in 2018. Her research focuses on Human-Computer Interaction and healthcare.
\end{IEEEbiography}
\vspace{-33pt}
\begin{IEEEbiography}[{\includegraphics[width=1in,height=1.25in,clip,keepaspectratio]{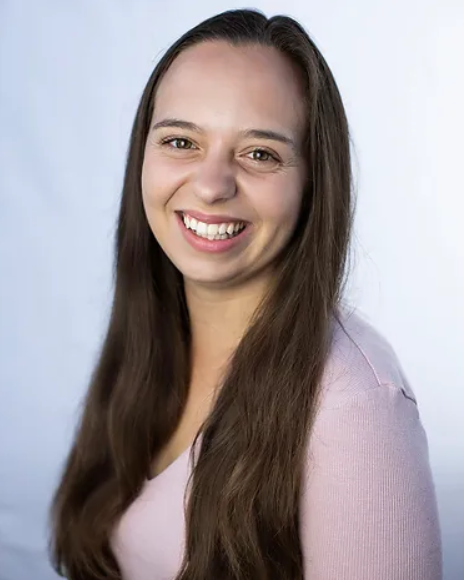}}]{Haley Foladare}
is a user researcher located in Vancouver, and research associate at the  UBC SPIN Lab, practicing a range of methodologies and streamlining research operations. She prioritizes an empathetic yet strategic approach that provides insights to designers and stakeholders while maintaining a focus on the needs and contexts of users.
\end{IEEEbiography}
\vspace{-33pt}
\begin{IEEEbiography}[{\includegraphics[width=1in,height=1.25in,clip,keepaspectratio]{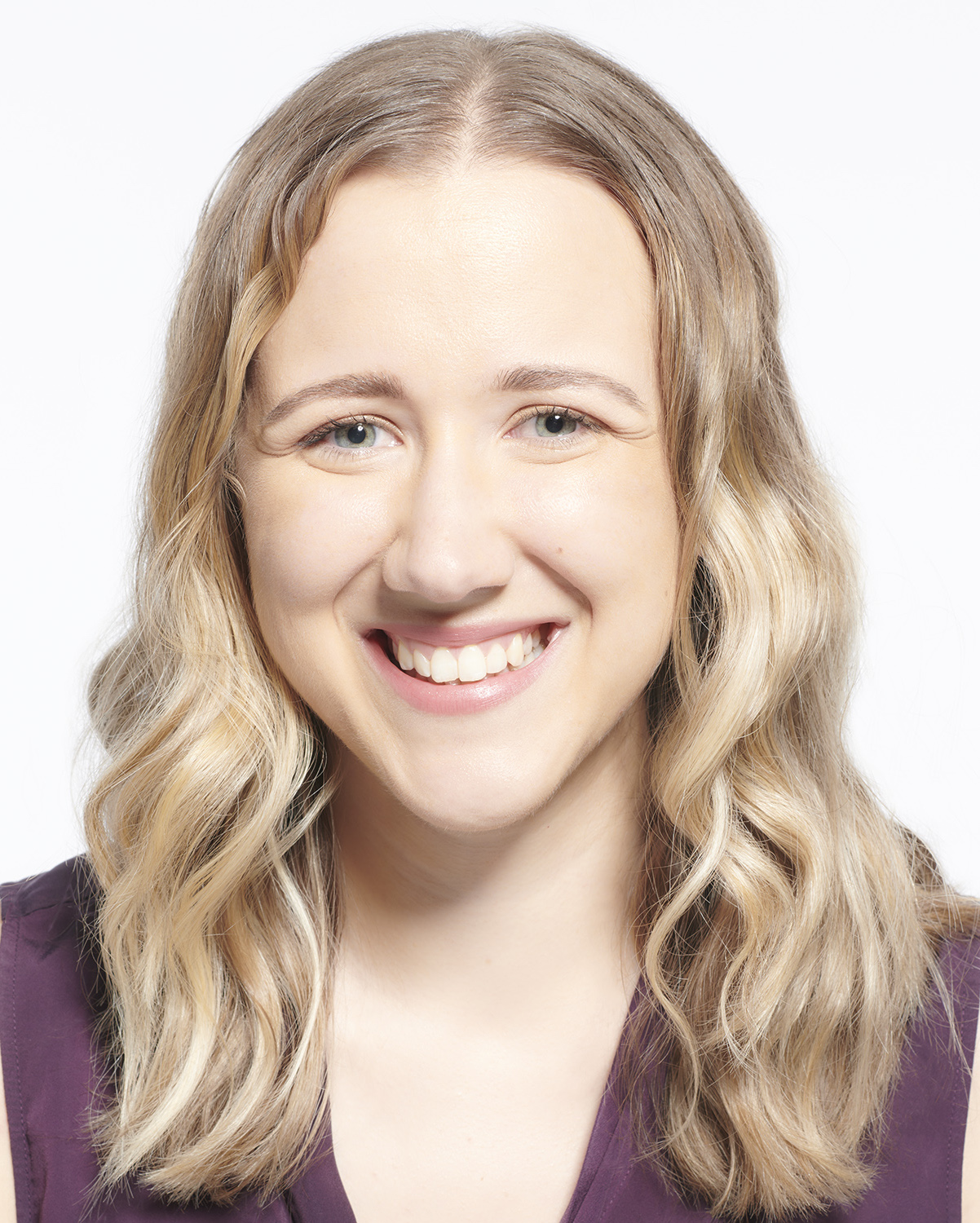}}]{Katelynn E. Boerner}
received her PhD in Clinical Psychology at Dalhousie University. She completed her residency in child clinical psychology at the Children’s Hospital of Eastern Ontario, and postdoctoral fellowships in Psychiatry and Pediatrics at the University of British Columbia. She works as a researcher studying health equity in pediatric pain, and a registered psychologist for the Complex Pain Service at BC Children’s Hospital.
\end{IEEEbiography}
\vspace{-33pt}
\begin{IEEEbiography}[{\includegraphics[width=1in,height=1.25in,clip,keepaspectratio]{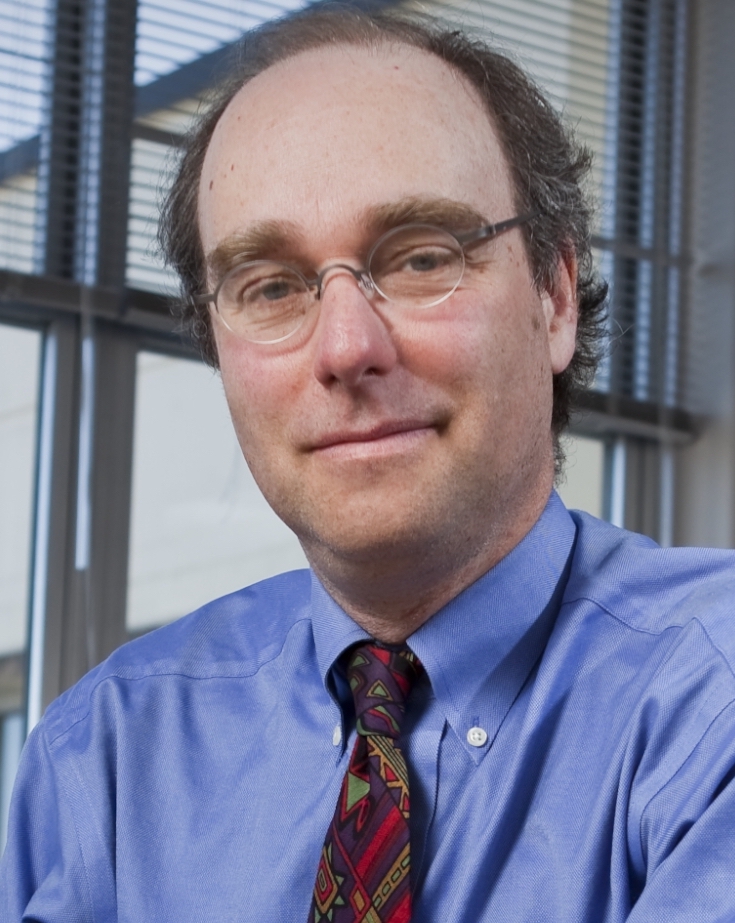}}]{Tim F. Oberlander}
is a physician-scientist  bridging developmental neuroscience and community child health. He is a developmental pediatrician 
studying how early social experience (prenatal maternal mental illness and psychotropic medication exposure) influences the developmental origins of stress and self-regulation and its impact on thinking, learning and behaviour during childhood. He is the medical lead for the Complex Pain Service at BC Children’s Hospital. 
\end{IEEEbiography}
\vspace{-33pt}
\begin{IEEEbiography}[{\includegraphics[width=1in,height=1.25in,clip,keepaspectratio]{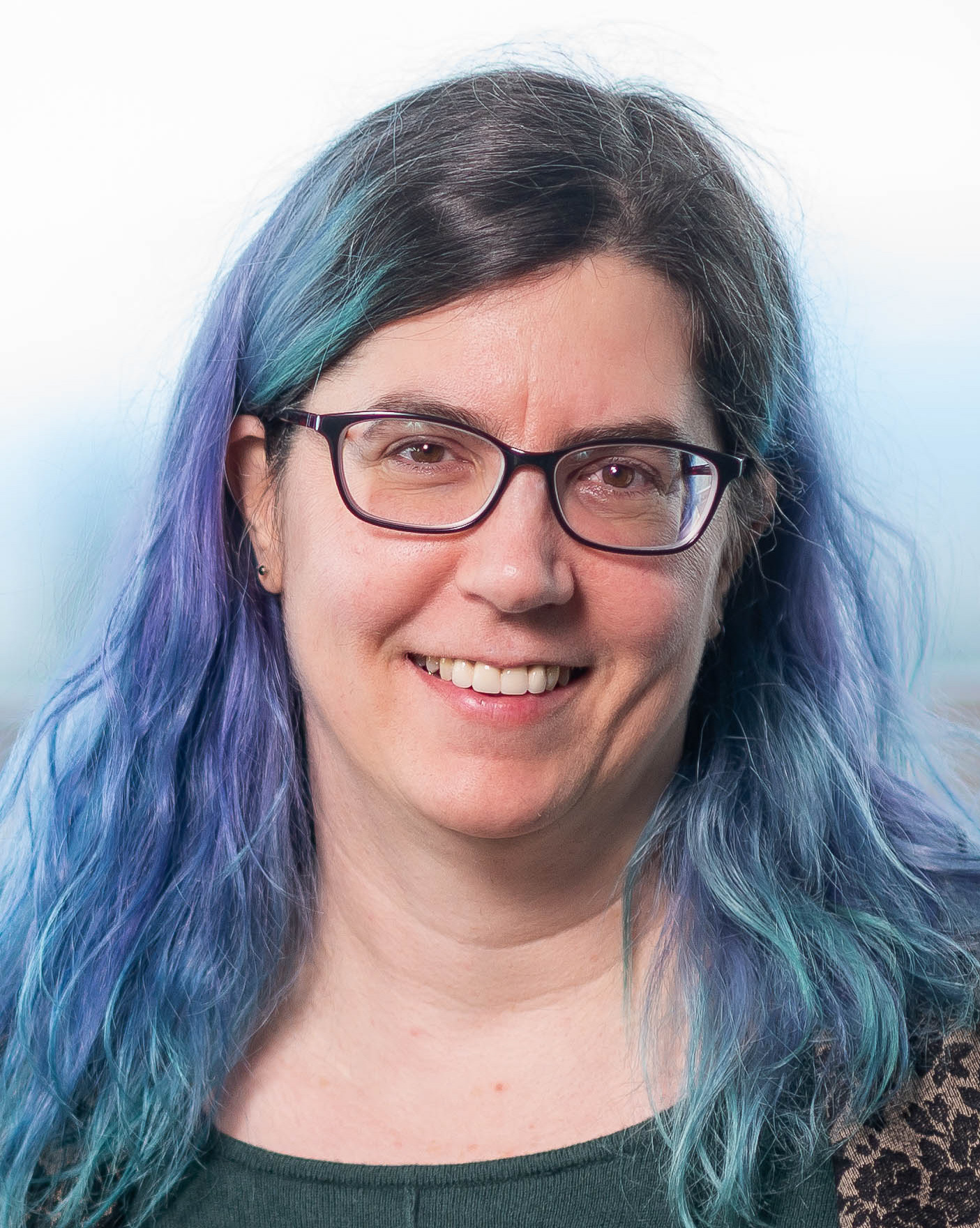}}]{Tamara Munzner}
(Fellow, IEEE) received the PhD degree from Stanford. She is currently a professor with the University of British Columbia. Her interests include the pursuit, methodology, and evaluation of visualization from both problem-driven and technique-driven perspectives. Her book Visualization Analysis and Design is heavily used worldwide, and she was the recipient of the IEEE VGTC Visualization Technical Achievement Award.
\end{IEEEbiography}
\vspace{-33pt}
\begin{IEEEbiography}[{\includegraphics[width=1in,height=1.25in,clip,keepaspectratio]{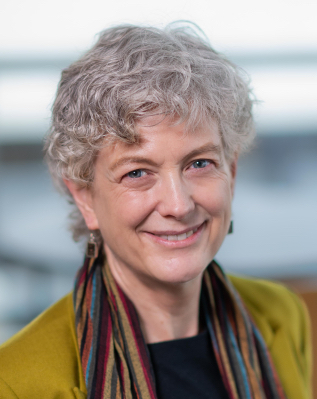}}]{Karon E. MacLean}
(Fellow, IEEE; Canada Research Chair in Interactive Human Systems Design) 
holds degrees in Biology and Mechanical Engineering (BSc, Stanford; M.Sc. / Ph.D, MIT).
She designs haptic and multimodal interactions, consideringcognition, perception and affect for both typical and special populations. She leads UBC's Designing for People interdisciplinary research cluster, and bridges haptics, HCI and health tech communities.
\end{IEEEbiography}

\vfill



\end{document}